\documentclass[twocolumn,epjc3]{svjour3} 
\RequirePackage{graphicx}
\RequirePackage{slashed}
\RequirePackage{booktabs,array}
\RequirePackage{bigints}
\RequirePackage{relsize}
\RequirePackage[b]{esvect}
\RequirePackage{caption}
\RequirePackage[colorlinks,citecolor=blue,urlcolor=blue,linkcolor=blue]{hyperref}

\newcommand{\diag}[3]{\raisebox{#3cm}{\includegraphics[width=#2cm]{figures/#1-eps-converted-to.pdf}}}
\newcommand\colequal{\mathrel{\overset{\makebox[0pt]{\mbox{\normalfont\tiny\sffamily col.}}}{=}}}
\newcommand\softequal{\mathrel{\overset{\makebox[0pt]{\mbox{\normalfont\tiny\sffamily soft}}}{=}}}

\begin{document}

\title{Collinear Electroweak Radiation in Antenna Parton Showers}

\author{Ronald Kleiss\thanksref{addr1}
        \and
        Rob Verheyen\thanksref{addr2,addr1}
}

\date{}

\institute{Radboud University Nijmegen, 6500 GL Nijmegen, The Netherlands \label{addr1}
           \and
           University College London, WC1E 6BT London, United Kingdom \label{addr2}
}

\maketitle 
\begin{abstract}
We present a first implementation of collinear electroweak radiation in the Vincia parton shower. 
Due to the chiral nature of the electroweak theory, explicit spin dependence in the shower algorithm is required.
We thus use the spinor-helicity formalism to compute helicity-dependent branching kernels, 
taking special care to deal with the gauge relics that may appear in computation that involve longitudinal polarizations of the massive electroweak vector bosons.
These kernels are used to construct a shower algorithm that includes all possible collinear final-state electroweak branchings, 
including those induced by the Yang-Mills triple vector boson coupling and all Higgs couplings, as well as vector boson emissions from the initial state. 
We incorporate a treatment of features particular to the electroweak theory, such as the effects of bosonic interference and recoiler effects, as well as a preliminary description of the overlap between electroweak branchings and resonance decays.
Some qualifying results on electroweak branching spectra at high energies, as well as effects on LHC physics are presented.
Possible future improvements are discussed, including treatment of soft and spin effects, as well as issues unique to the electroweak sector.
\end{abstract}

\section{Introduction}
Beyond the discovery of the Higgs boson \cite{ATLASHiggs,CMSHiggs}, signs of new physics have yet to appear at the LHC and the Standard Model has so far survived all forms of scrutiny. 
It has therefore become more likely that the Standard Model continues to describe nature accurately up to very high energy scales. 
At these very high energies heavy particles like electroweak gauge bosons, Higgs bosons and top quarks can start to appear as constituents of jets \cite{topInJets1,topInJets2} or otherwise contribute to radiative corrections. 
Virtual corrections have been shown to become sizeable even at LHC energies in exclusive observables \cite{EWLarge1,EWLarge2,EWLarge3,EWLarge4,EWLarge5,EWLarge6,EWLarge7,EWLarge8,EWLarge9,diJetLarge1,diJetLarge2,VjLarge1,VjLarge2,VjLarge3}. 
For instance, corrections to transverse momentum at LHC energies can already reach about 10\% for exclusive dijet production \cite{diJetLarge1,diJetLarge2}, 
and about 20\% for single vector boson production \cite{VjLarge1,VjLarge2,VjLarge3}, and they can be expected to grow even larger at future collider energies \cite{100TeV1,100TeV2}. 

Work such as \cite{SMFFs,EWInclusive} has instead focussed on the incorporation of electroweak logarithms in the resummation of inclusive observables at high energies.
However, many practical observables are not fully exclusive or fully inclusive.
In most cases, the only practical solution is to instead include EW effects in a systematic way in parton shower algorithms as part of a 
general-purpose event generator such as Pythia \cite{Pythia8.2}, Sherpa \cite{Sherpa} or Herwig \cite{Herwig++}.
These produce fully differential final states, and can be set up to include specific phenomena exclusive to the EW sector such the effects of symmetry breaking and gauge boson decay.
Furthermore, the inclusion of EW effects in parton shower algorithms enables automatic interleaving with the QCD shower, 
as well as the interfacing with the models of nonperturbative aspects of the event generation process, such as multi-parton interactions and hadronization.
As an example, ATLAS has reported on measurements that are sensitive to the collinear enhancements associated with $W$ radiation in jets \cite{ATLASEW}, 
and compared them to event generators that incorporate these effects.
There, it is also pointed out that these types of effects will play a significant role for several measurements at high energy scales, which will become more abundant as the LHC gathers more data. 

Electroweak corrections have been incorporated in parton showers in the past. 
An implementation \cite{PythiaEW,PythiaEW2} is available in Pythia event generator \cite{Pythia8.2} which only includes the radiation of electroweak gauge bosons and does not retain any spin information.
The radiation of electroweak gauge bosons was similarly included in the Sherpa event generator \cite{Sherpa} to study $W$ emissions in jets \cite{SherpaEW}. 
Another approach \cite{ALPGENEW} was employed in ALPGEN \cite{ALPGEN} where fixed-order matrix element calculations are combined with analytic Sudakov factors to achieve results similar to those of an electroweak parton shower.
A more recent work \cite{Tweedie} has implemented a final-state electroweak shower in the Pythia $1 \rightarrow 2$ transverse momentum ordered shower formalism that 
retains spin information and includes all branchings present in the electroweak sector.

In this paper, extend the antenna-based Vincia parton shower \cite{VinciaSimple,VinciaTimelike,VinciaHadron1} to include all possible final-state collinear EW branchings, 
including those associated with Yang-Mills triple vector boson and Higgs couplings. 
Furthermore, collinear vector boson emissions off the initial state are also included.
Vincia is a plugin to the Pythia event generator and already allows for QCD evolution with partons of definite helicity states \cite{VinciaHelicity1,VinciaHelicity2}.
This feature is especially important in the electroweak theory due to its chiral nature. 
The electroweak shower described in this paper will thus be responsible for the electroweak component of the shower evolution, and is interleaved with the default Vincia QCD shower.

The formalism descibed in this paper makes use of the spinor-helicity formalism to compute the large number of spin-dependent branching kernels associated with the electroweak shower.
Helicity-dependent QCD antenna functions have previously been computed with comparable methods \cite{Peskin1,Peskin2}.
We start with a brief overview of the spinor-helicity formalism and the conventions used in the calculation of the branching kernels. 
In section \ref{branchingAmplitudesSection}, the spinor-helicity formalism is used to compute branching kernels for all branching processes in the electroweak sector.
Section \ref{limitsSection} discusses the collinear limits of those branching kernels given in terms of Altarelli-Parisi splitting functions \cite{DGLAP3}.
The results in that section are found to be in agreement with \cite{Tweedie}.
Section \ref{implementationSection} details the implementation of an electroweak shower in the Vincia framework and treats a number of peculiarities exclusive to the electroweak sector 
such as the presence of bosonic interference and the matching to resonance decays.
To show the significance of an electroweak shower, its effects are investigated in section \ref{resultsSection} for highly energetic particles, but also at LHC energies.
We finally conclude in section \ref{conclusion}, discussing a number of missing features that are required to bring the modelling of EW radiation further in line with current QCD showers.

\section{The Spinor-Helicity Formalism} \label{SpinorHelicitySection}
Due to the chiral nature of the electroweak theory, it is important to calculate electroweak branching kernels for individual spin states. 
We choose to perform these calculations using the spinor-helicity formalism using conventions similar to those described in \cite{spinorHelicity}. 
The branching kernels computed here describe the correct soft and collinear factorization properties of branchings in the electroweak sector, while remaining Lorentz-invariant and independent of a particular representation of the Dirac algebra or explicit forms for fermionic spinors.
Furthermore, their analytic nature will be useful in dealing with issues that appear due to gauge dependence for longitudinal polarizations of gauge bosons.
We first briefly summarize our conventions and techniques.

\subsection{Spinors}
Helicity spinors for massive fermions may be defined as 
\begin{align} \label{standardFormMassive}
u_{\lambda}(p) &= \frac{1}{\sqrt{2 p{\cdot}k}}(\slashed{p} + m) u_{-\lambda}(k) \mbox{ and } \nonumber \\
v_{\lambda}(p) &= \frac{1}{\sqrt{2 p{\cdot}k}}(\slashed{p} - m) u_{\lambda}(k),
\end{align}
where $\lambda$ is the fermion helicity and $k$ is a lightlike reference vector that defines the meaning of the helicity of the fermion.
Due to its massive nature, helicity is not a Lorentz-invariant quantity and does not coincide with the chirality of the fermion. 
The spin vector associated with the spinors defined in eq.~\eqref{standardFormMassive} is 
\begin{equation} \label{massiveSpinVector}
s_{\lambda}^{\mu} = \frac{\lambda}{m} \left(p^{\mu} - \frac{m^2}{p{\cdot}k} k^{\mu} \right). 
\end{equation}
We therefore choose the reference vector 
\begin{equation} \label{refVector}
k = (1,-\vv{e}),
\end{equation}
where $\vv{e}$ is a unit vector pointing in the direction of $\vv{p}$.
With this choice, the massive helicity spinors retain the usual meaning of helicity as the projection of spin along the direction of motion.

\subsection{Polarization Vectors}
The polarization vectors for a massive vector boson with momentum $p$ are defined as
\begin{align} \label{standardFormPolarizations}
\epsilon_{\pm}^{\mu}(p) &= \pm \frac{1}{\sqrt{2}} \frac{1}{2 p{\cdot}k} \bar{u}_{\mp}(k) \slashed{p} \gamma^{\mu} u_{\pm}(k) \mbox{ and } \nonumber \\
\epsilon_{0}^{\mu}(p) &= \frac{1}{m} \left(p^{\mu} - 2 \frac{m^2}{2 p{\cdot}k} k^{\mu} \right),
\end{align}
where $k$ is again a massless reference vector.
Here, $\epsilon_{\pm}^{\mu}(p)$ are the transverse polarizations and $\epsilon_{0}^{\mu}(p)$ is the purely longitudinal polarization 
which only exists for massive vector bosons. 
By again choosing eq.~\eqref{refVector}, it is immediately clear that $\epsilon_{\pm}^{\mu}(p)$ are purely transverse and $\epsilon_{0}^{\mu}(p)$ is purely longitudinal.

\subsection{Spinor Products and Amplitude Evaluation}
Having expressed all massive spinors and polarization vectors in terms of massless spinors, 
amplitudes for particles with definite helicities can now be calculated very efficiently. 
We first define the spinor product 
\begin{equation} \label{spinorProduct}
S_{\lambda}(k_a, k_b) \equiv \bar{u}_{\lambda} (k_a) u_{-\lambda}(k_b)
\end{equation}
for lightlike (reference) vectors $k_a$ and $k_b$, which obey 
\begin{equation}
|S_{\lambda}(k_a, k_b)|^2 = 2 k_a{\cdot}k_b.
\end{equation} 
Spinor products are not uniquely defined, but one possible representation is given by
\begin{align} \label{spinorProductForm}
S_{\lambda}(k_a, k_b) &= (\lambda k_a^2 + i k_a^3) \sqrt{\frac{k_b^0 - k_b^1}{k_a^0 - k_a^1}} \nonumber \\
    &- (\lambda k_b^2 + i k_b^3)\sqrt{\frac{k_a^0 - k_a^1}{k_b^0 - k_b^1}}.
\end{align}
Using the spinors and polarization vectors of the previous section, all amplitudes can be expressed in terms of these spinor products. 
Structures like
\begin{equation} \label{spinorProductGeneral}
S_{\lambda}(k_a, p_i, p_j, ...,k_b) \equiv \bar{u}_{\lambda}(k_a) \slashed{p}_i \slashed{p}_j ... u_{\pm \lambda} (k_b),
\end{equation}
may still appear, where $p_i, p_j, ...$ may be massive. 
Such structures may be expressed in terms of the spinor products eq.~\eqref{spinorProduct} by defining
\begin{equation}
\hat{p}_i = p_i - \frac{p_i^2}{2 p_i{\cdot} k_i} k_i,
\end{equation}
which is explicitly massless. 
Eq.~\eqref{spinorProductGeneral} may then be written as
\begin{equation}
S_{\lambda}(k_a, p_i, p_j, ...,k_b) = S_{\lambda}(k_a, \hat{p}_i) \, S_{-\lambda}(\hat{p}_i, p_j,..., k_b).
\end{equation} 
This procedure is then repeated, the next time making $p_j$ massless by subtracting $(p_j^2 / 2 \hat{p}_i {\cdot}p_j) \hat{p}_i$, 
until only spinor products remain.

\section{Electroweak Branching Amplitudes} \label{branchingAmplitudesSection}
We now use the spinor-helicity formalism to compute branching amplitudes for the electroweak sector.
We first recount the phase space regions where radiative amplitudes factorize into a non-radiative amplitude and a radiative correction. 
For the moment, we restrict ourselves to splittings in the final state, where we denote the splitting momenta as $p_{ij} \rightarrow p_i + p_j$.
The parent momentum $p_{ij}$ may be considered to be off-shell by an amount that vanishes in the singular limits.

We first consider the quasi-collinear limit \cite{QuasiCollinear1,QuasiCollinear2,QuasiCollinear3}, 
which may be defined using the Sudakov decomposition
\begin{align} \label{SudakovDecomposition}
p_i &= z p_{ij} + \alpha_i n + q_{\perp} \nonumber \\
p_j &= (1-z) p_{ij} + \alpha_j n - q_{\perp}
\end{align} 
where $n$ is a lightlike reference vector usually taken to be the anti-collinear direction,
and $q_{\perp}$ is the spacelike transverse momentum with respect to $p_{ij}$ and $n$. 
These momenta satisfy $q_{\perp}{\cdot}p_{ij} = 0$ and $q_{\perp}{\cdot}n = 0$.
The variable $z$ is the collinear momentum fraction, and the parameters $\alpha_i$ and $\alpha_j$ 
are fixed by the conditions $p_i^2 = m_i^2$ and $p_j^2 = m_j^2$.
The off-shellness $Q^2$ is then given by
\begin{align} \label{QuasiCollinearOffshellness}
Q^2 &= (p_i + p_j)^2 - m^2_{ij} \nonumber \\
&= \frac{p_{\perp}^2}{z(1-z)} + \frac{m_i^2}{z} + \frac{m_j^2}{1-z} - m_{ij}^2,
\end{align} 
where $p_{\perp}^2 = -q_{\perp}^2$.
The singular factorization of the radiative amplitude occurs when this off-shellness becomes small with respect to the energy scale of the process.
In the massless case, the collinear region is thus defined by $p_{\perp} \rightarrow 0$.
This limit then generalizes to the quasi-collinear limit in the massive case, defined by
\begin{equation} \label{quasiCollinearPhaseSpace}
p_{\perp}, m_i, m_j \rightarrow 0 \text{ with fixed ratios } \frac{m_i}{p_{\perp}}, \frac{m_j}{p_{\perp}}.
\end{equation} 
In this limit, matrix elements symbolically factorize as
\begin{equation} \label{EWcollinearFactorization}
|M_{n+1}|^2 \colequal \frac{1}{Q^2} P(\lambda_{ij}, \lambda_{i}, \lambda_{j} , z) |M_n|^2,
\end{equation}
where $P(\lambda_{ij}, \lambda_{i}, \lambda_{j} , z)$ is the helicity-dependent Altarelli-Parisi splitting kernel \cite{DGLAP1,DGLAP2,DGLAP3}.

The second singular limit is the soft limit, where 
\begin{equation} \label{softPhaseSpace}
E_j \rightarrow m_j \mbox{ and } E_i \gg E_j
\end{equation}
or equivalent with $p_i$ and $p_j$ switched, with the soft particle being a gauge boson. 
The amplitude exhibits the usual eikonal factorization
\begin{equation} \label{EWsoftFactorization}
M_{n+1} \softequal M_n \times c \, \frac{2 p_i{\cdot}\epsilon_{\lambda_j}}{Q^2}  \delta_{\lambda_{ij} \lambda_i},
\end{equation}
where $c$ is some spin-dependent coupling. 

Using the spinor-helicity formalism, we compute $1 \rightarrow 2$ electroweak branchings kernels that capture these soft and quasi-collinear factorization properties. 
To that end, we compute a branching amplitude 
\begin{equation}
M^{\{ij\} \rightarrow i j}(\lambda_{ij}, \lambda_i, \lambda_j) \equiv \diag{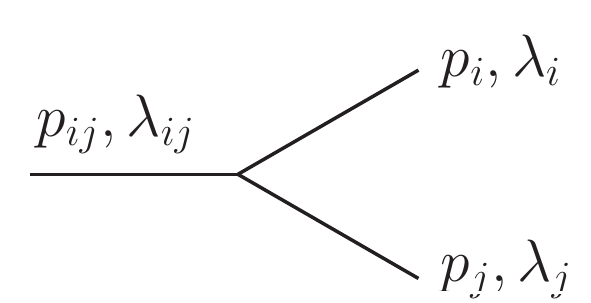}{3.4}{-0.65}
\end{equation} 
for every possible $1\rightarrow 2$ branching in the electroweak sector.
The branching kernel is then given by
\begin{align} \label{EWbranchingKernel}
B&_{\lambda_{ij}, \lambda_i, \lambda_j}(p_{ij}, p_i, p_j) = \frac{1}{Q^4} |M^{\{ij\} \rightarrow i j}(\lambda_{ij}, \lambda_i, \lambda_j)|^2.
\end{align} 
A comprehensive list of the branching amplitudes is given in \ref{EWbranchingAmplitudes}, which also includes vector boson emission from initial state fermions.
Here, we highlight one of these calculations to discuss the treatment of artifacts that may appear for longitudinal polarizations.

We consider vector boson emission from a fermion. 
The branching amplitudes are straightforwardly found to be 
\begin{align} \label{vectorBosonEmissionExample}
M&^{f\rightarrow f' V}(\lambda_{ij}, \lambda_i, \lambda_j) \nonumber \\
&= \bar{u}_{\lambda_i}(p_i) (v+a\gamma^5) \slashed{\epsilon}_{\lambda_j}(p_j)(\slashed{p}_{ij} + m_{ij}) u_{-\lambda_{ij}}(k_{ij}),
\end{align} 
where $v$ and $a$ are the vector and axial couplings of the fermion and vector boson at hand as defined in \ref{EWFeynRules}.
Note that this amplitude is applicable to all types of electroweak vector boson emission off fermion branching processes, including the cases where the fermion changes flavour due to a $W$-emission, 
or where the vector boson is massless and the axial coupling drops out in case of a photon.
For massive vector bosons, inserting the corresponding polarization vector leads to the appearance of terms proportional with 
\begin{align} \label{unitarityViolatingTerm}
\frac{1}{m_j} \slashed{p}_i \slashed{p}_j \slashed{p}_{ij} = \frac{1}{m_j} \left( (Q^2 + m_{ij}^2) \slashed{p}_i - m_i^2 \slashed{p}_{ij} \right).
\end{align}
The term proportional to $Q^2$ cancels the propagator, leaving a contribution that is not singular in the quasi-collinear or soft limit. 
We note that similar situations may appear in equivalent calculations of QCD branchings kernels. 
Nonsingular gauge-dependent terms may appear that do not contribute to the logarithmic accuracy of a parton shower \cite{NLL1,NLL2} and may be used as parameters for uncertainty estimation \cite{VinciaTimelike}.
In the case of emission of electroweak gauge bosons, the situation is however not the same. 
Due to the $\mathcal{O}(E/m)$ scaling of the longitudinal polarization, the leftover contributions lead to unphysical unitarity-violation at large energies. 

Such problematic terms originate from the scalar component of the longitudinal polarization of the electroweak gauge bosons, which originate from their corresponding Goldstone bosons.
In the above example, this Goldstone boson couples to the fermion through its Yukawa, which in this case appears as its kinematic off-shell mass. 
Due to the analytic nature of the spinor-helicity formalism, the offending terms are straightforwardly identified and corrected. 
Equivalently, the unitarity-violating term of eq.~\eqref{unitarityViolatingTerm} cannot survive if all other Feynman diagrams of a complete branching process are included, 
while all other terms are associated with a propagator factor $1/Q^2$ which cannot be cancelled elsewhere. 
In eq.~\eqref{vectorBosonEmissionExample} the unitarity-violating terms may thus be removed by the replacement 
\begin{equation}
\slashed{p}_i \slashed{p}_j \slashed{p}_{ij} \rightarrow m_{ij}^2 \slashed{p}_i - m_i^2 \slashed{p}_{ij}.
\end{equation} 
The replacements required for all other branching amplitudes are described in \ref{EWbranchingAmplitudes}.

In leading-colour QCD showers, soft gluon interference may be approximately incorporated by angular ordering \cite{QCDcolliderPhysics}, 
or by $2 \rightarrow 3$ dipole-like functions\cite{HerwigDipole,SherpaShower,DireShower} or antennae functions \cite{Ariadne,VinciaTimelike,VinciaHadron1}.
In the EW sector, no leading-colour approximation exists and many eikonal terms may contribute to the soft limit of an EW gauge boson emission.
An algorithm that includes the full multipole structure of QED radiation in the Vincia shower was described in \cite{QEDantenna,Multipole}. 
The inclusion of the all soft interference effects is accomplished by defining a $n \rightarrow n+1$ branching kernel and sectorizing the phase space into multiple regions to regulate the intricate soft and collinear singular structure.
An extension of this algorithm requires the computation of an equivalent spin-dependent $n \rightarrow n+1$ branching kernel for soft-enhanced EW gauge boson radiation,
as well as an implementation that involves a similar sectorization of phase space.
This is beyond the scope of this paper and is left for future work.

\section{Collinear Limits of the Branchings Amplitudes} \label{limitsSection}
In this section we discuss the behaviour of the branching amplitudes in the quasi-collinear limit described by eq.~\eqref{quasiCollinearPhaseSpace}. 
In this limit, the reference vectors simplify to
\begin{equation}
k_i \colequal k_j \colequal k_{ij} \equiv k.
\end{equation}
The branching amplitudes can be expressed in terms of the energy sharing variable $z$ by replacing 
\begin{equation}
p_i \rightarrow z \, p_{ij} \mbox{ and } p_j \rightarrow (1-z) \, p_{ij}.
\end{equation}
The only two remaining spinor products in the branching amplitudes are related by
\begin{equation}
S_{-\lambda}(k, p_j, p_i, k) = -S_{-\lambda}(k, p_i, p_j, k).
\end{equation}
Up to a phase factor, they are
\begin{align} \label{phaseFactor}
S_{-\lambda}&(k, p_i, p_j, k) \propto \nonumber \\
    & 2 \sqrt{p_i {\cdot} k_i \, p_j {\cdot} k_j \vphantom{Q^2 + m_{ij}^2 - m_i^2 \frac{p_{ij} {\cdot} k_{ij}}{p_i{\cdot}k_i} - m_j^2 \frac{p_{ij} {\cdot} k_{ij}}{2 p_j {\cdot} k_j}}} 
    \sqrt{Q^2 + m_{ij}^2 - m_i^2 \frac{p_{ij} {\cdot} k_{ij}}{p_i{\cdot}k_i} - m_j^2 \frac{p_{ij} {\cdot} k_{ij}}{p_j {\cdot} k_j}} \nonumber \\
& \colequal 2 p_{ij} {\cdot} k \sqrt{z(1-z) \vphantom{\tilde{Q}^2}} \sqrt{\tilde{Q}^2}
\end{align}
where 
\begin{equation}
\tilde{Q}^2 = Q^2 + m_{ij}^2 - \frac{m_j^2}{1-z} - \frac{m_i^2}{z}.
\end{equation}
Tables \ref{Amps1}-\ref{Amps7} contain the collinear limits of all electroweak branching amplitudes. 
These limits are related to the Altarelli-Parisi splitting kernels by
\begin{equation}
P(\lambda_{ij}, \lambda_{i}, \lambda_{j} , z) = |M(\lambda_{ij}, \lambda_{i}, \lambda_{j})|^2
\end{equation} 
where $M$ is the branching amplitude.
Note that we used the replacement of eq.~\eqref{phaseFactor} for the sake of notation, leading to a missing phase factor that is irrelevant for the computation of the splitting kernels.
The splitting functions found here agree with the results of \cite{Tweedie}.

\begin{table}
\bgroup
\def\arraystretch{1.7}
\begin{tabular}{lll|l}
    $\lambda_{ij}$ & $\lambda_i$ & $\lambda_j$ & $f \rightarrow f' V \mbox{ and } \bar{f} \rightarrow \bar{f}' V$ \\ \cline{1-4}
    $\lambda$ & $\lambda$ & $\lambda$ &  \scriptsize $\sqrt{2} \lambda (v-\lambda a) \sqrt{\tilde{Q}^2} \frac{1}{\sqrt{1-z}} $ \normalsize \\
    $\lambda$ & $\lambda$ & $-\lambda$ & \scriptsize $\sqrt{2} \lambda (v-\lambda a) \sqrt{\tilde{Q}^2} \frac{z}{\sqrt{1-z}} $ \normalsize \\
    $\lambda$ & $-\lambda$ & $\lambda$ & \scriptsize $\sqrt{2} \lambda  \bigg[ m_{ij} (v - \lambda a) \sqrt{z} - m_i (v+\lambda a) \frac{1}{\sqrt{z}} \bigg]$ \normalsize \\
    $\lambda$ & $-\lambda$ & $-\lambda$ & \scriptsize 0 \normalsize \\
    $\lambda$ & $\lambda$ & $0$ & \scriptsize $\begin{aligned} &(v-\lambda a) \bigg[\frac{m_{ij}^2}{m_j} \sqrt{z} - \frac{m_i^2}{m_j} \frac{1}{\sqrt{z}} - 2 m_j \frac{\sqrt{z}}{1-z} \bigg] \\
        &+ (v+\lambda a) \frac{m_i m_{ij}}{m_j} \frac{1-z}{\sqrt{z}} \end{aligned}$ \normalsize \\
    $\lambda$ & $-\lambda$ & $0$ & \scriptsize $ \sqrt{\tilde{Q}^2} \sqrt{1-z \vphantom{\tilde{Q^2}}} \bigg[ \frac{m_i}{m_j} (v-\lambda a) - \frac{m_{ij}}{m_j} (v+\lambda a) \bigg] $ \normalsize
\end{tabular}
\egroup
\caption{Branching amplitudes for vector boson emission off a fermion. 
For the antifermion, the interchange $(v-\lambda a) \leftrightarrow (v+\lambda a)$ is applied.}
\label{Amps1}
\end{table}

\begin{table}
\bgroup
\def\arraystretch{1.7}
\begin{tabular}{lll|l}
    $\lambda_I$ & $\lambda_i$ & $\lambda_j$ & $V \rightarrow f \bar{f}'$ \\ \cline{1-4}
    $\lambda$ & $\lambda$ & $-\lambda$ & \scriptsize $\sqrt{2} \lambda (v-\lambda a) \sqrt{\tilde{Q}^2} \, z $ \normalsize \\
    $\lambda$ & $-\lambda$ & $\lambda$ & \scriptsize $\sqrt{2} \lambda (v+\lambda a) \sqrt{\tilde{Q}^2} \, (1-z)$ \normalsize \\
    $\lambda$ & $\lambda$ & $\lambda$ & \scriptsize $\sqrt{2} \lambda \bigg[ m_i (v + \lambda a) \sqrt{\frac{1-z}{z}} + m_j (v - \lambda a) \sqrt{\frac{z}{1-z}} \bigg]$ \normalsize \\
    $\lambda$ & $-\lambda$ & $-\lambda$ & \scriptsize 0 \normalsize \\
    $0$ & $\lambda$ & $\lambda$ & \scriptsize $\sqrt{\tilde{Q}^2} \bigg[ \frac{m_i}{m_{ij}} (v+\lambda a) + \frac{m_j}{m_{ij}} (v-\lambda a) \bigg] $ \normalsize \\
    $0$ & $\lambda$ & $-\lambda$ & \scriptsize $\begin{aligned} &(v-\lambda a) \bigg[2 m_{ij} \sqrt{z(1-z)}  - \frac{m_i^2}{m_{ij}} \sqrt{\frac{1-z}{z}} \\ &-\frac{m_j^2}{m_{ij}} \sqrt{\frac{z}{1-z}}\bigg] 
    + (v+\lambda a) \frac{m_i m_j}{m_{ij}} \frac{1}{\sqrt{z(1-z)}} \end{aligned}$ \normalsize 
\end{tabular}
\egroup
\caption{Branching amplitudes for vector boson splitting to fermions.}
\label{Amps2}
\end{table}

\begin{table}
\bgroup
\def\arraystretch{1.7}
\begin{tabular}{lll|l}
    $\lambda_I$ & $\lambda_i$ & $\lambda_j$ & $V \rightarrow V' V'' \times g_{V}$ \\ \cmidrule(r{6.8cm}){1-4}
    $\lambda$ & $\lambda$ & $\lambda$ & \scriptsize $ \sqrt{2} \lambda \sqrt{\tilde{Q}^2} \sqrt{\frac{1}{z(1-z)} \vphantom{\tilde{Q^2}} }$ \normalsize \\
    $\lambda$ & $\lambda$ & $-\lambda$ & \scriptsize  $ \sqrt{2} \lambda \sqrt{\tilde{Q}^2}  z \sqrt{\frac{z}{1-z} \vphantom{\tilde{Q^2}} } $ \normalsize \\
    $\lambda$ & $-\lambda$ & $\lambda$ & \scriptsize  $ \sqrt{2} \lambda \sqrt{\tilde{Q}^2} (1-z) \sqrt{\frac{1-z}{z} \vphantom{\tilde{Q^2}} }$ \normalsize \\
    $\lambda$ & $-\lambda$ & $-\lambda$ & \scriptsize $ 0 $ \normalsize \\
    $0$ & $\lambda$ & $\lambda$ & \scriptsize 0 \normalsize \\
    $0$ & $\lambda$ & $-\lambda$ & \scriptsize $m_{ij} (2z - 1) + \frac{m_j^2}{m_{ij}} - \frac{m_i^2}{m_{ij}}$ \normalsize \\
    $\lambda$ & $0$ & $\lambda$ & \scriptsize $m_i \left(1 + 2 \frac{1-z}{z}\right) + \frac{m_j^2}{m_i} - \frac{m_{ij}^2}{m_i}$ \normalsize \\
    $\lambda$ & $0$ & $-\lambda$ &\scriptsize  0 \normalsize \\
    $\lambda$ & $\lambda$ & $0$ & \scriptsize $m_j \left(1 + 2 \frac{z}{1-z}\right) + \frac{m_i^2}{m_j} - \frac{m_{ij}^2}{m_j}$ \normalsize \\
    $\lambda$ & $-\lambda$ & $0$ & \scriptsize 0 \normalsize \\
    $\lambda$ & $0$ & $0$ & \scriptsize $\frac{\lambda}{\sqrt{2}} \frac{m_i^2 +m_j^2 - m_{ij}^2 }{m_i m_j} \sqrt{\tilde{Q}^2} \sqrt{z(1-z) \vphantom{\tilde{Q^2}}}$ \normalsize \\
    $0$ & $\lambda$ & $0$ & \scriptsize $\frac{\lambda}{\sqrt{2}} \frac{m_{ij}^2 + m_j^2 - m_i^2}{m_{ij} m_j} \sqrt{\tilde{Q}^2} \sqrt{\frac{1-z}{z} \vphantom{\tilde{Q^2}}}$ \normalsize \\ 
    $0$ & $0$ & $\lambda$ & \scriptsize $\frac{\lambda}{\sqrt{2}} \frac{m_{ij}^2 + m_i^2 - m_j^2}{m_{ij} m_i} \sqrt{\tilde{Q}^2} \sqrt{\frac{z}{1-z} \vphantom{\tilde{Q^2}}}$ \normalsize \\
    $0$ & $0$ & $0$ & \scriptsize $\begin{aligned}  &\frac{1}{2} \frac{m_{ij}^3}{m_i m_j} (2z-1) - \frac{m_i^3}{m_{ij} m_j} \left( \frac{1}{2} + \frac{1-z}{z} \right) \\
    &+ \frac{m_j^3}{m_{ij} m_i} \left(\frac{1}{2} + \frac{z}{1-z} \right) + \frac{m_i m_j}{m_{ij}} \left( \frac{1-z}{z} - \frac{z}{1-z} \right) \\
    &+ \frac{m_{ij} m_i}{m_j} (1-z) \left(2 + \frac{1-z}{z} \right) - \frac{m_{ij} m_j}{m_i} z \left(2 + \frac{z}{1-z} \right) \end{aligned}$ \normalsize
\end{tabular}
\egroup
\caption{Branching amplitudes for vector boson emission off a vector boson.}
\label{Amps3}
\end{table}

\begin{table}
\bgroup
\def\arraystretch{1.7}
\begin{tabular}[t]{ll|l}
    $\lambda_I$ & $\lambda_i$ & $(f \rightarrow f h \mbox{ and } \bar{f} \rightarrow \bar{f} h) \times \frac{e}{2 s_w} \frac{m_f}{m_w} $ \\ \cline{1-3}
    $\lambda$ & $\lambda$ & \scriptsize $m_f \Big[ \sqrt{z} + \frac{1}{\sqrt{z}} \Big]$ \normalsize \\
    $\lambda$ & $-\lambda$ & \scriptsize $\sqrt{1-z \vphantom{\tilde{Q^2}}} \sqrt{\tilde{Q}^2}$\normalsize 
\end{tabular}
\egroup
\caption{Branching amplitudes for Higgs emission off (anti)fermions.}
\label{Amps4}
\end{table}

\begin{table}
\bgroup
\def\arraystretch{1.7}
\begin{tabular}{ll|l}
    $\lambda_i$ & $\lambda_j$ & $h \rightarrow f \bar{f} \times \frac{e}{2 s_w} \frac{m_f}{m_w} $ \\ \cline{1-3}
    $\lambda$ & $\lambda$ & \scriptsize $\sqrt{\tilde{Q}^2}$  \normalsize \\
    $\lambda$ & $-\lambda$ & \scriptsize $m_f \Big[ \sqrt{\frac{1-z}{z} \vphantom{\tilde{Q^2}}} - \sqrt{\frac{z}{1-z} \vphantom{\tilde{Q^2}}} \Big]$ \normalsize 
\end{tabular}
\egroup
\caption{Branching amplitudes for Higgs splitting to fermions.}
\label{Amps5}
\end{table}

\begin{table}
\bgroup
\def\arraystretch{1.7}
\begin{tabular}{ll|l}
    $\lambda_I$ & $\lambda_i$ & $V \rightarrow V h \times g_h $ \\ \cline{1-3}
    $\lambda$ & $\lambda$ &  $-1$ \\
    $\lambda$ & $-\lambda$ & $0$ \\ 
    $0$ & $\lambda$ & $\frac{1}{m_{ij}}  \frac{\lambda}{\sqrt{2}} \sqrt{\tilde{Q}^2} \sqrt{z(1-z) \vphantom{\tilde{Q^2}}} $ \\
    $\lambda$ & $0$ & $\frac{1}{m_i} \frac{\lambda}{\sqrt{2}} \sqrt{\tilde{Q}^2} \sqrt{\frac{1-z}{z} \vphantom{\tilde{Q^2}}}$ \\
    $0$ & $0$ & $\frac{1}{2} \frac{m_j^2}{m_i^2} + \frac{1-z}{z} + z$
\end{tabular}
\egroup
\caption{Branching amplitudes for Higgs emission off a vector boson.}
\label{Amps6}
\end{table}

\begin{table}
\bgroup
\def\arraystretch{1.7}
\begin{tabular}{ll|l}
    $\lambda_i$ & $\lambda_i$ & $h \rightarrow V V \times g_V$ \\ \cline{1-3}
    $\lambda$ & $\lambda$ &  $0$ \\
    $\lambda$ & $-\lambda$ & $-1$ \\ 
    $0$ & $\lambda$ & $\frac{1}{m_i} \frac{\lambda}{\sqrt{2}} \sqrt{\tilde{Q}^2} \sqrt{\frac{1-z}{z} \vphantom{\tilde{Q^2}}} $ \\
    $\lambda$ & $0$ & $\frac{1}{m_j}  \frac{\lambda}{\sqrt{2}} \sqrt{\tilde{Q}^2} \sqrt{\frac{z}{1-z} \vphantom{\tilde{Q^2}}} $ \\
    $0$ & $0$ & $\frac{1}{2} \frac{m_{ij}^2}{m_i^2} - 1 - \frac{1-z}{z} - \frac{z}{1-z}$
\end{tabular}
\egroup
\caption{Branching amplitudes for Higgs splitting to vector bosons.}
\label{Amps7}
\end{table}
The quasi-collinear limits of the branching amplitudes listed in Tables \ref{Amps1}-\ref{Amps7} show a rich landscape of splitting modes that are the result of several physical effects. 
For instance, in the case of vector boson emission off a fermion, listed in Table \ref{Amps1}, we find that the splitting functions for a transverse emission without a spin flip correspond with the familliar spin-summed form 
\begin{equation} \label{vectorSplittingFunction}
P_{f \rightarrow f'V} \propto \frac{1 + z^2}{1-z},
\end{equation}
in correspondence with the QCD equivalent of gluon emission off a quark. 
The presence of fermion and vector boson masses induces a shift of $1/Q^2 \rightarrow \tilde{Q}^2 / Q^4$ in the propagator structure. 
The fermionic mass corrections in $\tilde{Q}^2$ also appear for gluon emission and reproduce the mass contributions to the eikonal factor in the soft limit. 
For $W$ and $Z$ emission, a vector boson mass correction is also present.
In case of a transverse emission with a spin flip, the splitting functions are suppressed with a relative factor $m^2/Q^2$, where $m$ may the the vector boson mass or either of the fermion masses.
Again, similar mass corrections appear in QCD \cite{HerwigSpin}, and cause such modes to be suppressed at off-shellness $Q^2$ much larger than the electroweak scale. 

The splitting function for longitudinal vector boson emission with a spin flip behaves like the scalar splitting function
\begin{equation} \label{scalarSplittingFunction}
P_{f \rightarrow f\varphi} \propto (1-z).
\end{equation}
In the unbroken Standard Model, this splitting mode indeed corresponds with the emission of a Goldstone boson, and the proportionality constant is given by the Yukawa coupling to the Goldstone. 
The spin flip mode is in this case not mass suppressed, reflecting the behaviour of scalar emissions in the unbroken phase. 
On the other hand, the mode without spin flip is mass suppressed, and terms that scale as $1/m_j$ appear, originating from the scalar piece of the longitudinal polarization, as well as terms that scale as $m_j$, corresponding with the vector piece.

\section{The Electroweak Shower Implementation} \label{implementationSection}
We use the branching kernels computed in the previous section to implement an electroweak shower in the Vincia framework, which was set out in \cite{VinciaSimple,VinciaTimelike,VinciaMassive,VinciaHelicity1,VinciaHelicity2,VinciaHadron1,VinciaHadron2,VinciaResonance}.
Here we first provide a brief summary before continuing with a description of some details specific to the electroweak shower.

Parton showers are constructed as a Markov chain of emissions that are distributed according to an approximation to the radiative matrix element and an associated Sudakov factor \cite{Review1}. 
In most modern showers, these branchings are kinematically modelled as $2 \rightarrow 3$ processes while the dynamics vary according to the shower model.
If both branching partons are in the final state, momenta are labelled as $I, K \rightarrow i,j,k$, and the Vincia shower approximation to the matrix element may be written as 
\begin{equation} \label{FFPartonShowerApproximation}
|M_{n+1}|^2 d\Phi_{n+1} = |M_n|^2 d\Phi_{n} \times a(s_{ij}, s_{jk}) \, d\Phi_{\mbox{\tiny{ant}}},
\end{equation} 
where 
\begin{equation} \label{FFantPhaseSpace}
d\Phi_{\mbox{\tiny{ant}}}^{\mbox{\tiny{FF}}} = \frac{1}{16 \pi^2} m_{IK}^2 \lambda^{-\frac{1}{2}}(m_{IK}^2, m_I^2, m_K^2) ds_{ij} ds_{jk} \frac{d\varphi}{2\pi},
\end{equation} 
where $s_{ij} = 2 p_i{\cdot}p_j$ and $\lambda$ is the K{\"a}ll{\'e}n function.
Eq.~\eqref{FFantPhaseSpace} represents an exact factorization of the radiative phase space. 
An associated kinematic map is defined between the pre-branching and post-branching momenta that conserves total momentum and is soft- and collinear-safe \cite{VinciaMassive}.
The branching kernels are so-called antenna functions $a(s_{ij}, s_{jk})$ that capture the leading collinear and soft singularities associated with QCD emissions.
The equivalent expressions to eq.~\eqref{FFPartonShowerApproximation} for radiation from the initial state, as well as the definition of the kinematic maps, may be found in \cite{VinciaHadron2}.
In the electroweak sector, we instead use the branching kernels computed in Section \ref{branchingAmplitudesSection}.
These kernels contain the correct singular behaviour in the soft-collinear and collinear limits, and are simultaneously usable for the treatment of bosonic interference effects and recoiler selection to be discussed below.

Vincia supports QCD evolution of partons with definite helicity \cite{VinciaHelicity1,VinciaHelicity2}, making for a natural framework for the inclusion of an electroweak shower. 
Interference effects between intermediate spin configurations have previously been incorporated in parton showers, such as in the Herwig \cite{Herwig++,Herwig7} parton shower \cite{HerwigSpin} and in Deductor \cite{NagyInterference2,NagyInterference3}.
The Vincia parton shower currently makes no attempt to incorporate such interferences, and correspondingly the same is true for the electroweak sector.

\subsection{Ordering and Resonance Showers} \label{orderingSubsection}
The electroweak shower includes a number of branchings that would normally be associated with the decay of resonances by Pythia \cite{Pythia6.4}, for which Vincia is a plugin.
In particular, the Standard Model particles that have such decay-like branchings are $Z$, $W^{\pm}$, Higgs and top quark. 
With the inclusion of an electroweak shower, the decay modes of the resonances are now also all present as shower branchings. 
The shower enables highly-energetic resonances to branch and disappear at scales much higher than their width, where they should indeed be treated as any other non-resonant particle. 
At scales close to the resonance width, the Breit-Wigner character of the resonance decay should instead dominate the distribution. 

Here, we set up a method of matching the parton shower distribution to a Breit-Wigner smoothly. 
However, a full-fledged incorporation of resonance decays at high scales in the Pythia event generator framework is a much more involved issue.
In the absence of EW shower branchings, Pythia always associates the scale of a resonance decay with the width of the resonance.
As the widths of all SM resonances are close to $\Lambda_{\mbox{\tiny{QCD}}}$, the approach of Pythia is to factorize the QCD shower of the hard scattering from showers in resonance decay systems.
When resonances are instead able to decay at much larger scales, this approximation is no longer valid and a more general solution is required.

Most generally, resonances should be allowed to decay at any scale during the shower evolution of the hard scattering. 
Once a decay occurs, a factorized shower in the resonance system may be performed using Vincia's resonance shower described in \cite{VinciaResonance}.
Afterwards, the showered decay products may be joined with the rest of the hard system and the evolution can be continued.
As these modifications require substantal restructuring of both the Pythia and Vincia framework, they will be treated separately in \cite{UpcomingResonance}.

Here, we do however proceed to set out the matching of the EW shower to the Breit-Wigner shape through a simple sampling procedure.
To that end, we must first choose a suitable ordering scale.
An issue specific to resonance branchings is that there are regions of phase space where the off-shellness $Q^2$ is negative. 
As more negative values of off-shellness should correspond with shorter-lived resonances, we are led to define an ordering scale 
\begin{equation} \label{orderingScale}
|Q^2| = |s_{ij} + m_i^2 + m_j^2 - m_{ij}^2|,
\end{equation}
and equivalent for initial state branchings.
For branchings that are not of the resonant decay type, $Q^2$ is strictly positive and the ordering scale corresponds to off-shellness ordering, 
which indeed regulates the soft-collinear and collinear singularities.

Note that, even without the presence of EW masses, the choice of ordering scale is by no means unique and many shower algorithm employ different functional forms.
The inclusion of EW scale masses leads to further ambiguities, since the virtuality or transverse momentum of the branching is not the only scale considered to be small in the quasi-collinear limit.
In \cite{PythiaEW} the effects of shifting the ordering scale to account for the presence of gauge boson masses were found to lead to essentially identical results, motivating eq.~\eqref{orderingScale} as it leads to simpler matching to the Breit-Wigner shape.

For most types of non-resonant electroweak branchings the phase space is naturally cut off due to the masses of the post-branching momenta.
Beyond resonance branchings, photon emission is the only remaining branching that is not cut off naturally. 
They are instead cut off at the same scale as the QCD shower approximately equal to $\Lambda_{\mbox{\tiny{QCD}}}$, and QED radiation at lower scales is included as is described in \cite{Multipole}.

We set up the Breit-Wigner distribution by making the replacement 
\begin{equation}
\frac{1}{Q^4} \rightarrow \frac{1}{Q^4 + m^2 \Gamma^2},
\end{equation} 
in the relevant branching kernels computed in Section \ref{branchingAmplitudesSection}, where $\Gamma$ is the width of the resonance. 
These kernels are then normalized to represent a probability distribution from which off-shellness scales, kinematics and post-branching spin states may be sampled.
Because the ordering scale of the shower is given by eq.~\eqref{orderingScale}, it is straightforward to define a matching scale $|Q^2_{\mbox{\tiny{Match}}}|$ where the shower is matched to the Breit-Wigner. 
However, while it may be possible to pick the matching scale to ensure the distribution is continuous, it will in general not be smooth. 
Furthermore, the shape of the shower distribution at scales close to the resonance width depends on the starting scale and the antenna mass, so such a choice would in any case not always be continuous. 
We thus ensure the distribution is smooth by sampling the value of the matching scale for every resonance from the distribution 
\begin{equation} \label{matchingScaleDistribution}
\mathcal{P}(|Q^2_{\mbox{\tiny{Match}}}|) \propto \frac{m^2 \Gamma^2 |Q_{\mbox{\tiny{Match}}}^2|}{(Q_{\mbox{\tiny{Match}}}^4 + m^2 \Gamma^2)^2}.
\end{equation}
The shower contribution to the off-shellness spectrum is then effectively multiplied by a factor 
\begin{equation} \label{matchingScaleDistributionIntegral}
\bigintssss_0^{|Q^2|} \! \! \! \! \! \! \! \! d|Q^2_{\mbox{\tiny{Match}}}| \, \mathcal{P}(|Q^2_{\mbox{\tiny{Match}}}|) = \frac{Q^4}{Q^4 + m^2 \Gamma^2},
\end{equation}
which ensures a smooth suppression of the shower kernel. 
In this treatment, the total distribution is dominated by the shower at hight scales, while the Breit-Wigner dominates at low scales. 

The implementation within the parton shower framework is relatively straightforward. 
When a shower of a resonance is initiated, a matching scale $|Q^2_{\mbox{\tiny{Match}}}|$ is sampled from eq.~\eqref{matchingScaleDistribution}, serving as a local cutoff scale.
If during showering the branching scale drops below the matching scale, a new off-shellness scale is instead drawn from the Breit-Wigner distribution.
We emphasize that this solution serves as an approximate means of matching the shower to a Breit-Wigner, and a more sophisticated method that closely matches that of Pythia will be developed in \cite{UpcomingResonance}.

\subsection{Recoiler Selection} \label{spectatorSelectionSubsection}
While in the QCD portion of the Vincia shower the colour structure dictates the pairings of branching partons $I$ and $K$,
no such guidance exists in the electroweak sector.
Furthermore, the branching kernels only describe the soft-collinear and collinear singularities associated with the branching of particle $I$.
The choice of recoiler $K$ does not contribute to the accuracy of the shower in these limits.
It may however be chosen probabilistically to minimize the physical consequences of the recoil on previously generated branchings.
For the branching of particle $i$, the probability to select a spectator $j$ from a pool of $N$ available ones is
\begin{equation} \label{ariFac}
\mathcal{P}_{ij} = \frac{\left| M^{x \rightarrow i j} \right|^2}{\sum_{j'=1}^N \left| M^{x \rightarrow i j'} \right|^2}.
\end{equation}
That is, all candidate spectators $j$ have a probability to be assigned as a recoiler for particle $i$ if the pair $i,j$ may have been produced by an electroweak branching.
All of the contributions in the denominator of eq.~\eqref{ariFac} thus correspond with possible shower histories that contribute to the current state.
The selection of a spectator is then more likely if the shower history where the current brancher and that spectator were created by a previous branching. 

We clarify the choice of eq.~\eqref{ariFac} through an example, which may be expressed in terms of diagrams as 
\begin{align}
\Bigg\rvert \diag{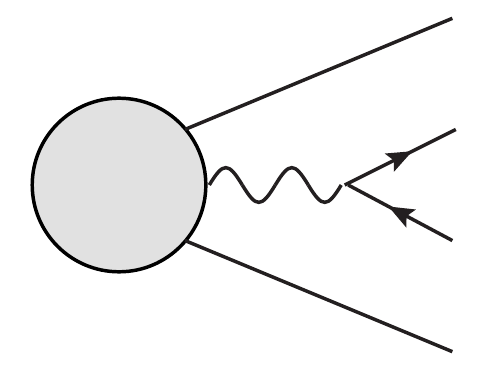}{1.5}{-0.5} \Bigg\rvert^2  
&= \frac{ \Big\rvert \diag{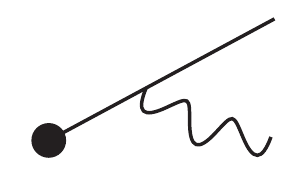}{1.2}{-0.3} \Big\rvert^2 }
{ \Big\rvert \diag{EW/ariFacEquation/frac1}{1.2}{-0.3} \Big\rvert^2 + \Big\rvert \diag{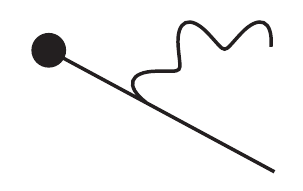}{1.4}{-0.35} \Big\rvert^2 }
\Bigg\rvert \diag{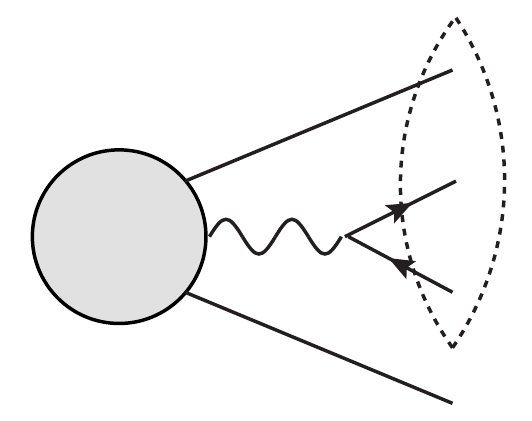}{1.6}{-0.46} \Bigg\rvert^2 \nonumber \\
&+\frac{ \Big\rvert \diag{EW/ariFacEquation/frac2}{1.2}{-0.35} \Big\rvert^2 }
{ \Big\rvert \diag{EW/ariFacEquation/frac1}{1.2}{-0.3} \Big\rvert^2 + \Big\rvert \diag{EW/ariFacEquation/frac2}{1.4}{-0.35} \Big\rvert^2 }
\Bigg\rvert \diag{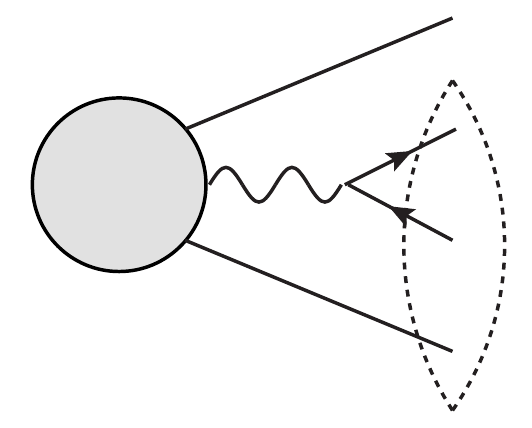}{1.6}{-0.62} \Bigg\rvert^2.
\end{align}
The spectator for the splitting of the vector boson is chosen to be either of the other external legs based on the probabilities that the vector boson was emitted by either of those legs. 
When the vector boson splits, it is brought off its mass shell by transferring some momentum of the spectator to the vector boson. 
Because the vector boson momentum is modified, the emission kernel that it was produced with is no longer correct. 
In the strong-ordering phase space region where $Q^2_{\mbox{\tiny{emit}}} \gg Q^2_{\mbox{\tiny{split}}}$ this type of mismodelling is absent.
However, the shower covers all of phase space, including regions where the scales $Q^2_{\mbox{\tiny{emit}}} > Q^2_{\mbox{\tiny{split}}}$ are of comparable size. 
Recoil effects of previous brancings may be especially relevant for branchings that involve masses of the order of the electroweak scale. 

The Vincia $2 \rightarrow 3$ kinematic map conserves the invariant mass of the original two-particle system.
This means that the probability of eq.~\eqref{ariFac} ensures the propagator structure of the emitter and vector boson pair that was most important in the emission process is most often conserved.
Note that a similar procedure has in the past been used to select a recoiler for gluon splitting \cite{Ariadne,VinciaMassive}, which also features only collinear singularities.
In \cite{Tweedie} this effect is referred to as `kinematic back-reactions` and is accounted for as a multiplicative factor of the branching kernels. 

\subsection{Bosonic Interference} \label{bosonicInterferenceSubsection}
A unique type of interference effect appears in the electroweak sector due to the existence of multiple neutrally charged bosons \cite{bosonicInterference}.
It is possible to treat such interference effects comprehensively by evolving density matrices that contain mixed states of neutral bosons \cite{Tweedie}, but such procedures quickly become computationally prohibitive.
We instead opt for a simpler approach that incorporates the most imporant physical effects while preventing the shower to become a bottleneck in the Monte Carlo event generation chain \cite{Computation}.

Interference between neutral bosons occurs when the electroweak shower produces a neutral boson, and it subsequently disappears by splitting. 
The heavy neutral bosons will always split due to the matching to a Breit-Wigner distribution described in Section \ref{orderingSubsection}, but photons may survive the showering procedure.
As such, interference effects are corrected for by applying an event weight 
\begin{equation} \label{bosonInterferenceCorrectionFactor}
w_{\mbox{\tiny{BI}}} = \sum_x \frac{|M^{x \rightarrow x' b_1} M^{b_1\rightarrow i j} + M^{x \rightarrow x' b_2} M^{b_2\rightarrow i j}|^2}{|M^{x \rightarrow x' b_1} M^{b_1\rightarrow i j}|^2 + |M^{x \rightarrow x' b_2} M^{b_2\rightarrow i j}|^2}
\end{equation} 
after the splitting, where we have dropped the helicity indices for readability.
Eq.~\eqref{bosonInterferenceCorrectionFactor} corrects the branching kernels for the interference between two bosons $b_1$ and $b_2$.
These bosons may be either a photon and a transversely polarized $Z$ boson, or a Higgs boson and a longitudinally polarized $Z$ boson. 
Interference between spin states is not taken into account in correspondence with the rest of the shower algorithm.
The weight of eq.~\eqref{bosonInterferenceCorrectionFactor} then sums over all possible emitters $x$ of the bosons $b_1$ and $b_2$, accounting for all possible shower histories.
This includes the different spin states of the same particle, but note that these contributions are summed over incoherently. 
The weight has the property $0 \leq w_{\mbox{\tiny{BI}}} \leq 2$, and since the overal rate of electroweak boson emissions is moderate even at high energies, there is little danger of wildly fluctuating weights leading to inefficiencies.

\subsection{Overestimate Determination} \label{overestimateDeterminationSubsection}
The implementation of electroweak radiation in the shower formalism through the Sudakov veto algorithm \cite{Fooling,Reloaded,Competing} requires finding overestimates of the associated branching kernels. 
Due to the large number of types of branchings in the electroweak sector, it is desirable to automate this procedure. 

For final-state particles with an electroweak charge, a recoiler is selected through the procedure outlined in Section \ref{spectatorSelectionSubsection}. 
The branching phase space is then given by eq.~\eqref{FFantPhaseSpace}.
All final-state electroweak branching kernels are overestimated by a parameterized function
\begin{align} \label{FFoverestimate}
\mathcal{O}^{\mbox{\tiny{FF}}} &= c_1^{\mbox{\tiny{FF}}} \frac{1}{|Q^2|} + c_2^{\mbox{\tiny{FF}}} \frac{1}{|Q^2|} \frac{E_{IK} (E_{IK} + |\vv{p}_{IK}|)}{s_{ij} + s_{ik} + m_i^2} \nonumber \\ 
&+ c_3^{\mbox{\tiny{FF}}} \frac{1}{|Q^2|} \frac{E_{IK} (E_{IK} + |\vv{p}_{IK}|)}{s_{ij} + s_{jk} + m_j^2} + c_4^{\mbox{\tiny{FF}}} \frac{m_I^2}{Q^4},
\end{align}
where $|Q^2|$ is given by eq.~\eqref{orderingScale}.
The term multiplying $c_1^{\mbox{\tiny{FF}}}$ reflects the $1/|Q^2|$ behaviour of most branching kernels, while the second and third terms incorporate the soft behaviour associated with vector boson emission, containing ratios $E_{IK}/E_i \sim 1/z$ and $E_{IK}/E_j \sim 1/(1-z)$ in terms of shower variables.
The term multiplying $c_4^{\mbox{\tiny{FF}}}$ represents the mass corrections that may be present for massive branchers. 
The contribution of post-branching masses are typically negative, and therefore do not improve the overestimate much.

Initial-state branchings are only allowed to recoil against other initial states. 
In this case, the antenna phase space is
\begin{equation} \label{IIantPhaseSpace}
d\Phi_{\mbox{\tiny{ant}}}^{\mbox{\tiny{II}}} = \frac{1}{16\pi^2} \frac{x_A^2}{x_a^2} \frac{x_B^2}{x_b^2} \frac{1}{s_{AB}} ds_{aj} ds_{bj} \frac{d\varphi}{2\pi},
\end{equation}
where $A$ branches to $a$ and $j$, and $B$ is the recoiler.
The electroweak shower currently only implements vector boson emission from fermions in the initial state, which are treated as massless by Vincia. 
The ordering scale is crossed into the initial state to give 
\begin{equation}
Q^2 = s_{aj} - m_j^2.
\end{equation}
An absolute value qualification is not required here since resonance type branchings do not occur in the initial state.
A sufficient overestimate is 
\begin{align}\label{IIoverestimate}
\mathcal{O}^{\mbox{\tiny{II}}} &= c_1^{\mbox{\tiny{II}}} \frac{1}{Q^2}\frac{s_{ab}}{s_{AB}} \nonumber \\
&+ c_2^{\mbox{\tiny{II}}} \frac{1}{Q^2} \frac{x_A^2 s_{ab}^2}{x_A s_{bj}(s_{ab} - s_{bj}) + x_B s_{aj}(s_{ab} - s_{aj})}.
\end{align}
The factor $s_{ab}/s_{AB}$ accounts for the additional factor of $1/z$ that shows up in the Altarelli-Parisi splitting kernels when crossed to the initial state. 
The second term represents the soft enhancement $1/(1-z)$ that may appear for vector boson emissions. 

The parameters $c_1^{\mbox{\tiny{FF}}}$ through $c_4^{\mbox{\tiny{FF}}}$, $c_1^{\mbox{\tiny{II}}}$ and $c_2^{\mbox{\tiny{II}}}$ are automatically determined for all possible branchings in the electroweak shower.
To do that, brancher-recoiler pairs are generated from antennae with randomly chosen invariant masses. 
Branchings are then generated with a distribution $1/|Q^2|$ for the final state or $s_{ab}/s_{AB} \, 1/Q^2$ for the initial state to roughly model the branching kernel behaviour. 
For every event $i$, the value of the branching kernel $B_i$ as well as the terms $A_{ij}$ multiplying the parameters $c_j$ are stored. 
The problem of finding suitable values for the overestimate parameters can then be formulated as 
\begin{align} \label{linearProgramming}
&\mbox{Minimize } \sum_{i=1}^n (A\mathbf{c})_i - B_i \nonumber \\
&\mbox{subject to } (A\mathbf{c})_i \geq B_i  \mbox{ and } \mathbf{c} \geq 0. 
\end{align}
The minimization condition minimizes the average difference between the branching kernel and its overestimate. 
The constraints ensure the overestimate is larger than the branching kernel for all samples and the parameters are positive definite. 
Eq.~\eqref{linearProgramming} is an instance of a linear programming problem, for which many libraries are available. 
We make use of the Python \cite{Python} package PuLP \cite{PuLP}. 

\subsection{Overview of the Shower Algorithm}
We conclude this section with a short description of the complete shower algorithm.
Branching kernels are constructed using the formalism described in section \ref{branchingAmplitudesSection} for all possible electroweak branchings and all helicity configurations.
Overestimates are found using the optimization algorithm of subsection \ref{overestimateDeterminationSubsection}, where the post-branching helicities are summed over.
This leaves a total of $277$ types of final-state branchings, of which $74$ are resonance decays, and $90$ types of initial-state branchings. 

As the shower initializes, a recoiler is selected for all final-state particles that have electroweak charge, making use of the selection probability described in subsection \ref{spectatorSelectionSubsection}.
Initial-state branchers are always paired with the other initial-state particle, recoiling against the entire event. 

While the shower runs, electroweak branchings compete against the QCD branchings generated by Vincia. 
The overestimates are used to generate trial branchings which are accepted or rejected through the usual Sudakov veto algorithm.
For resonance decay branchings, the procedure outlined in subsection \ref{orderingSubsection} is used to match the shower to a Breit-Wigner distribution. 
We make use of the same kinematic maps as Vincia, and first-order running of the electroweak coupling constant is incorporated as part of the veto procedure.
Wherever applicable, the bosonic interference weight eq.~\eqref{bosonInterferenceCorrectionFactor} is included. 
After accepting a branching, a helicity state is selected with probability
\begin{equation}
\mathcal{P}_{\lambda_I, \lambda_i, \lambda_j} = \frac{B_{\lambda_I, \lambda_i, \lambda_j}(p_I, p_i, p_j)}{\sum_{\lambda_i,\lambda_j}B_{\lambda_I, \lambda_i, \lambda_j}(p_I, p_i, p_j)},
\end{equation} 
and equivalent for resonance decay branchings.
The QCD shower and the electroweak shower run interleaved until the QCD cutoff scale is reached, after which only QED radiation is simulated. 

\section{Results} \label{resultsSection}
The inclusion of electroweak radiation in parton showers opens up a rich field of showering phenomena, in particular at energies well above the electroweak scale. 
Rather than considering specific sets of observables for specific processes, in this section we first consider radiation spectra of several particles at energies compatible with future colliders, 
showing overal branching rates of the electroweak shower, the relative importance of the large number of emission modes and the effects of bosonic interference.
We conclude by showing results for hadron collision processes at LHC energies and comparing with the Pythia electroweak shower \cite{PythiaEW}.

\subsection{Branching Spectra}
\begin{figure*} 
\centering
\begin{minipage}{.5\textwidth}
    \centering
    \includegraphics[scale=0.4]{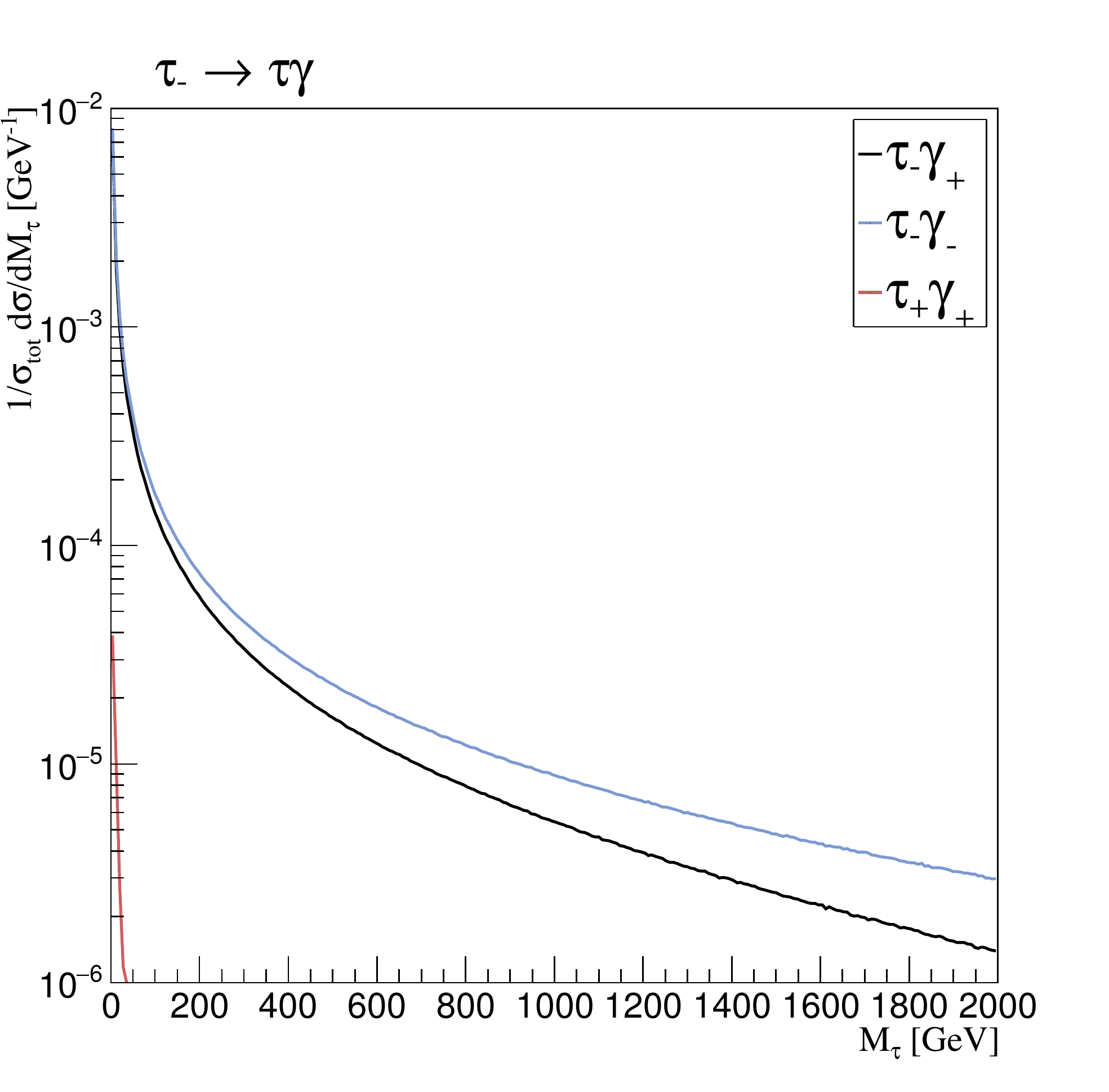}
\end{minipage}%
\begin{minipage}{.5\textwidth} 
    \centering
    \includegraphics[scale=0.4]{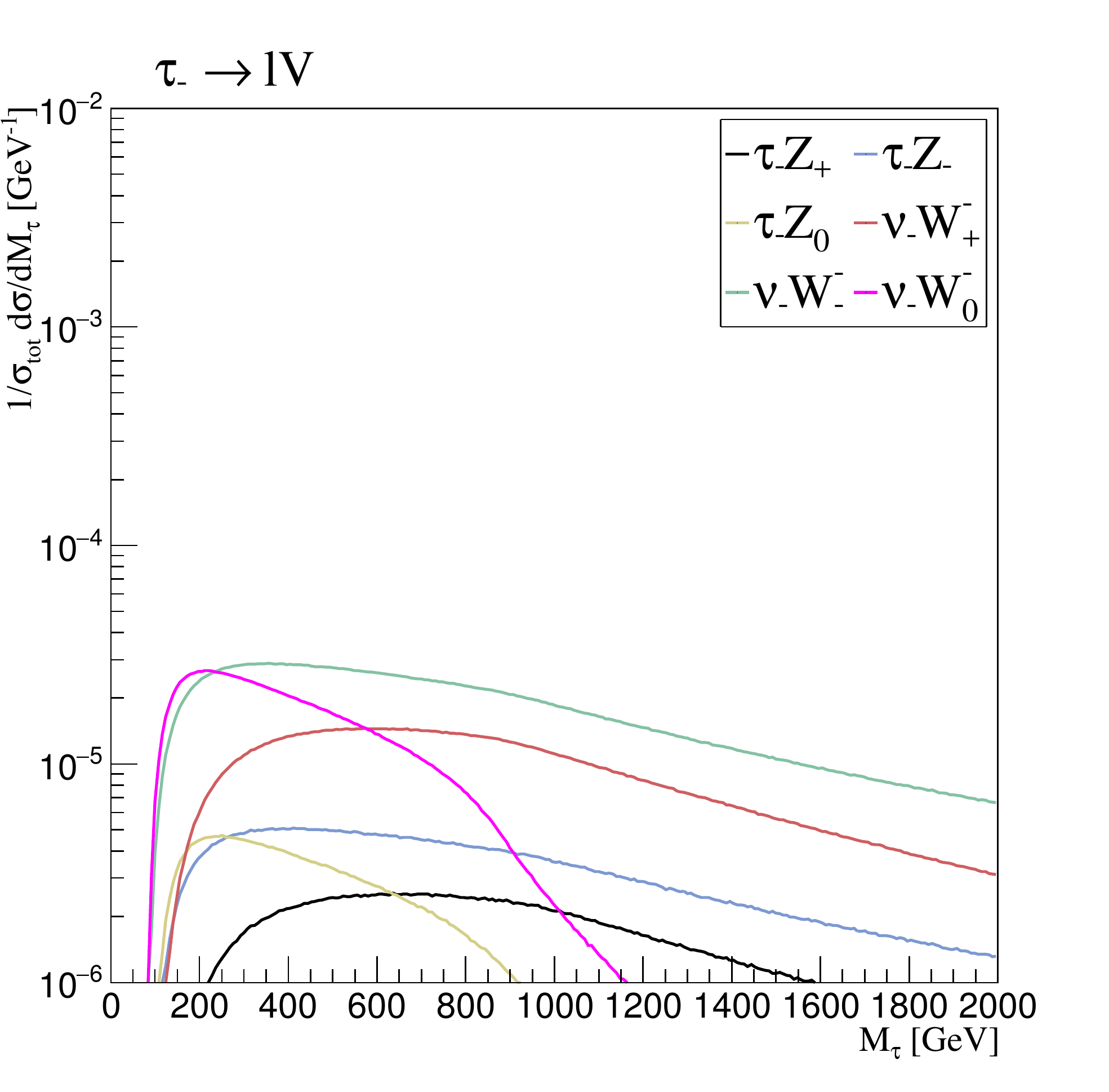}
\end{minipage}
\caption{Branching spectra of a $1$ TeV $\tau_{-}$ to $\tau \gamma$ (left) and $\tau Z$ and $\nu_{\tau} W^{-}$ (right).}
\label{tauEWshowerFigure}
\end{figure*}

\begin{figure*}
\centering
\begin{minipage}{.5\textwidth}
    \includegraphics[scale=0.4]{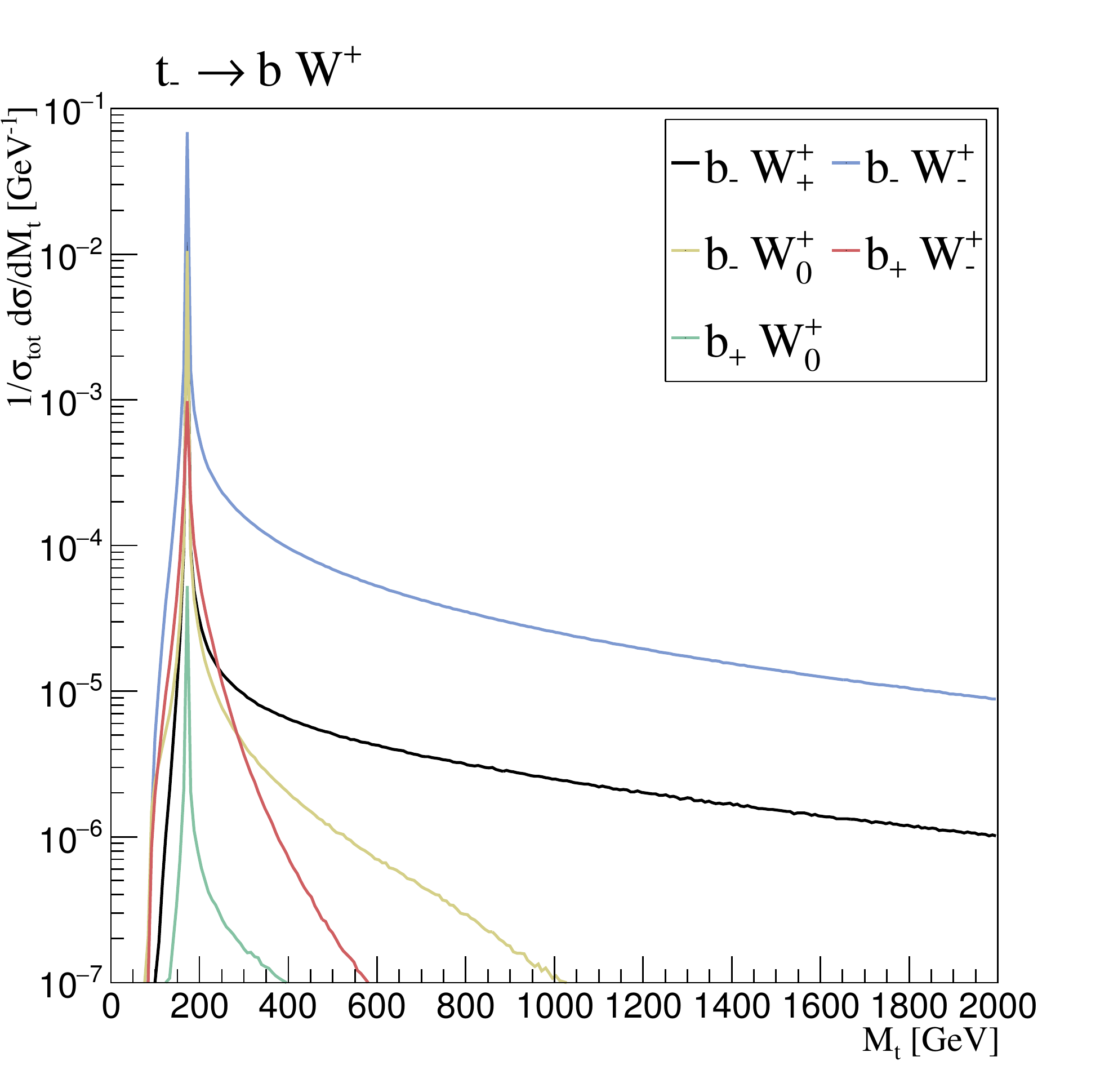}
\end{minipage}%
\begin{minipage}{.5\textwidth} 
    \includegraphics[scale=0.4]{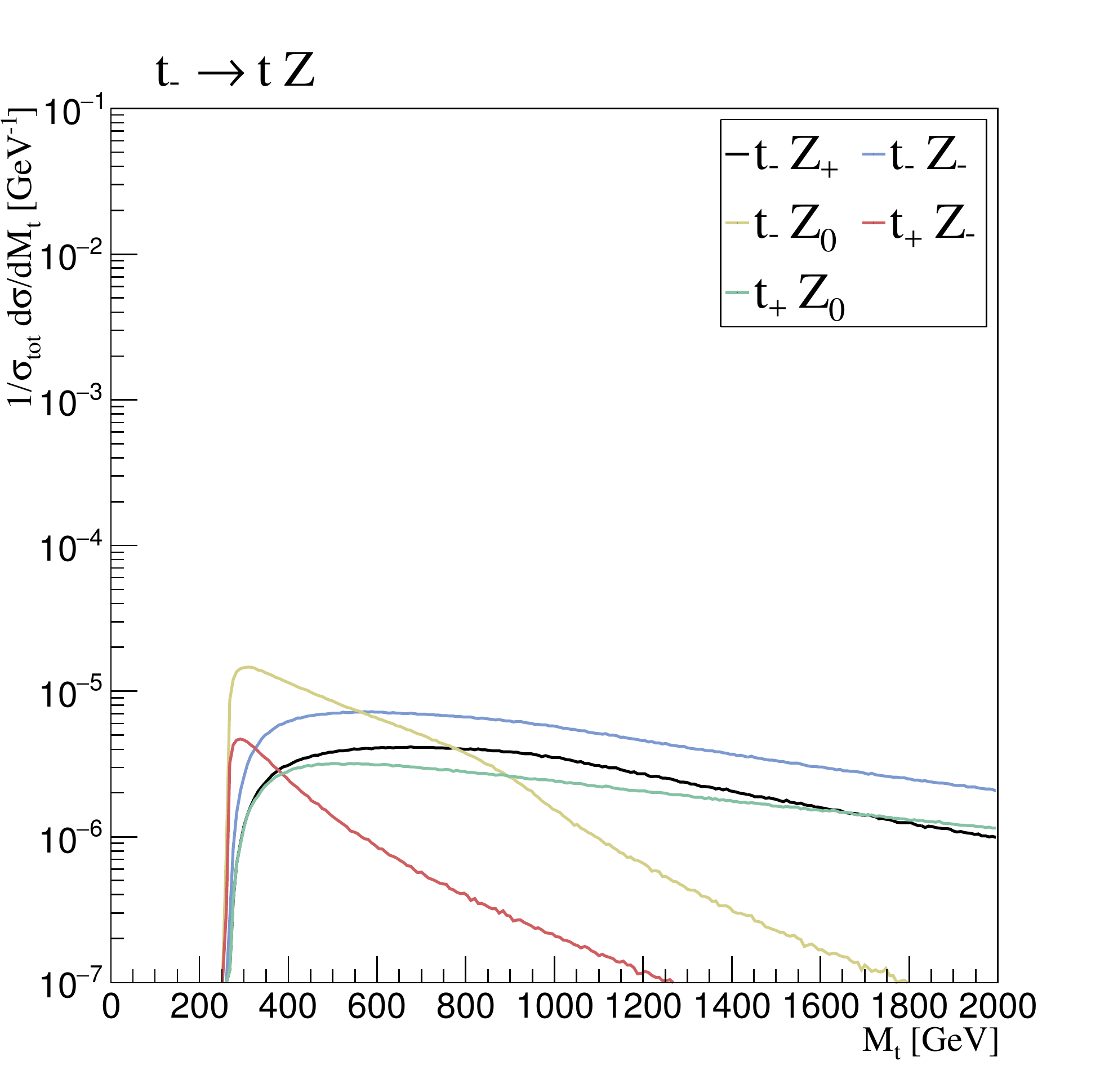}
\end{minipage}
\includegraphics[scale=0.4]{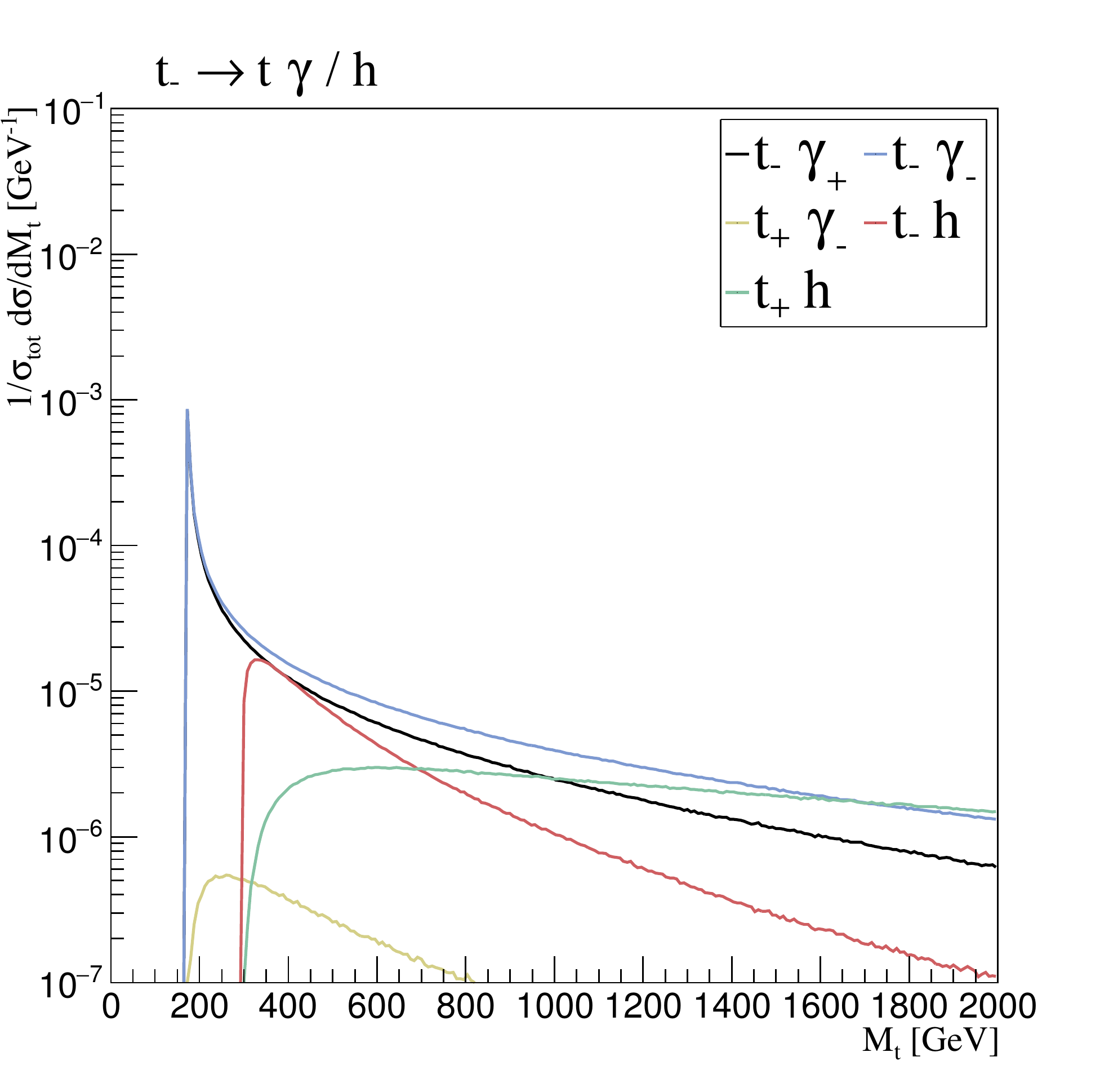}
\caption{Branching spectra of a $1$ TeV top quark to $b W^+$ (left), $tZ$ (right) and $t \gamma/h$ (bottom).}
\label{topEWshowerFigure}
\end{figure*}

\begin{figure*} 
\centering
\begin{minipage}{.5\textwidth}
    \includegraphics[scale=0.4]{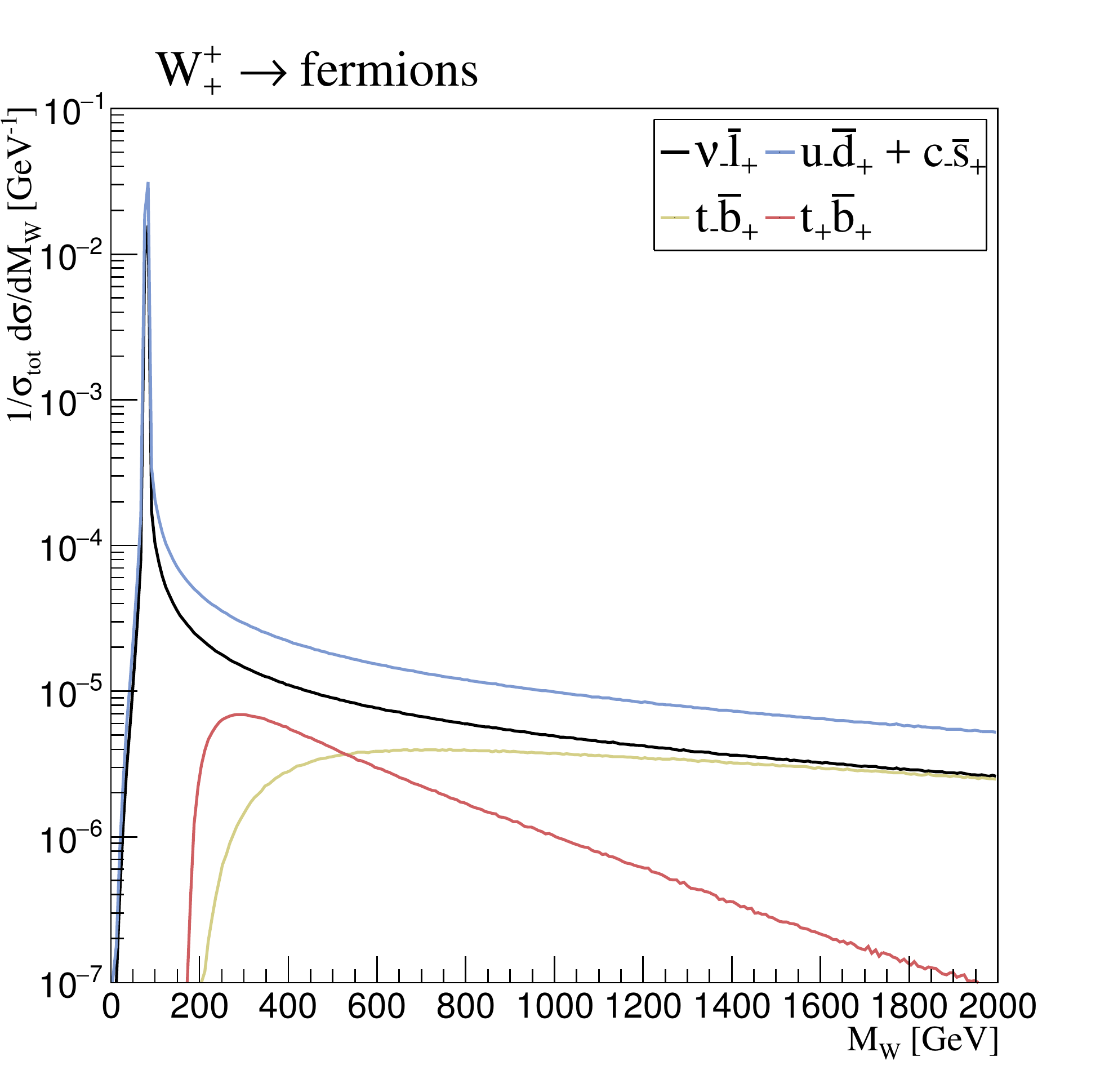}
\end{minipage}%
\begin{minipage}{.5\textwidth} 
    \includegraphics[scale=0.4]{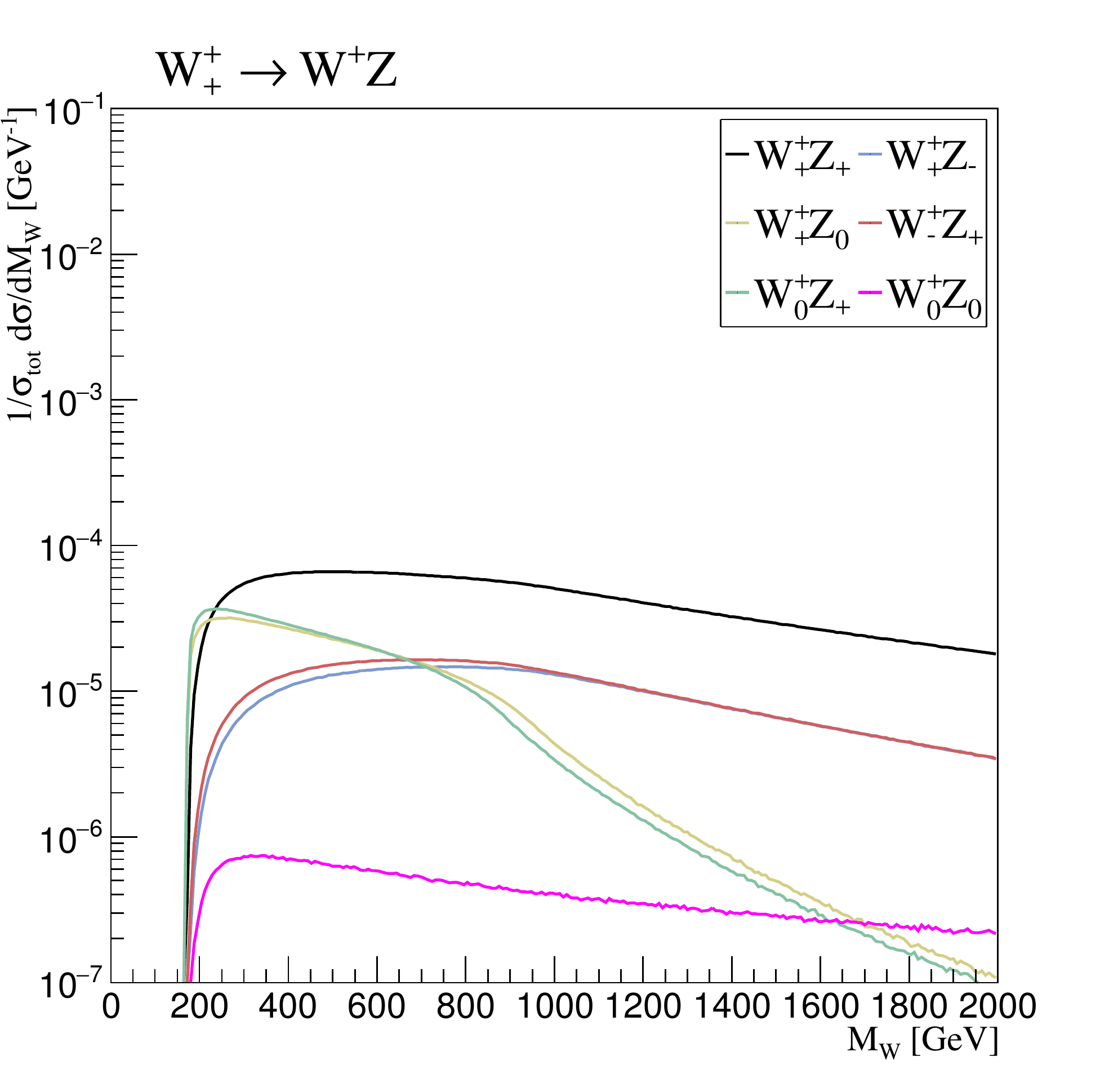}
\end{minipage}
\includegraphics[scale=0.4]{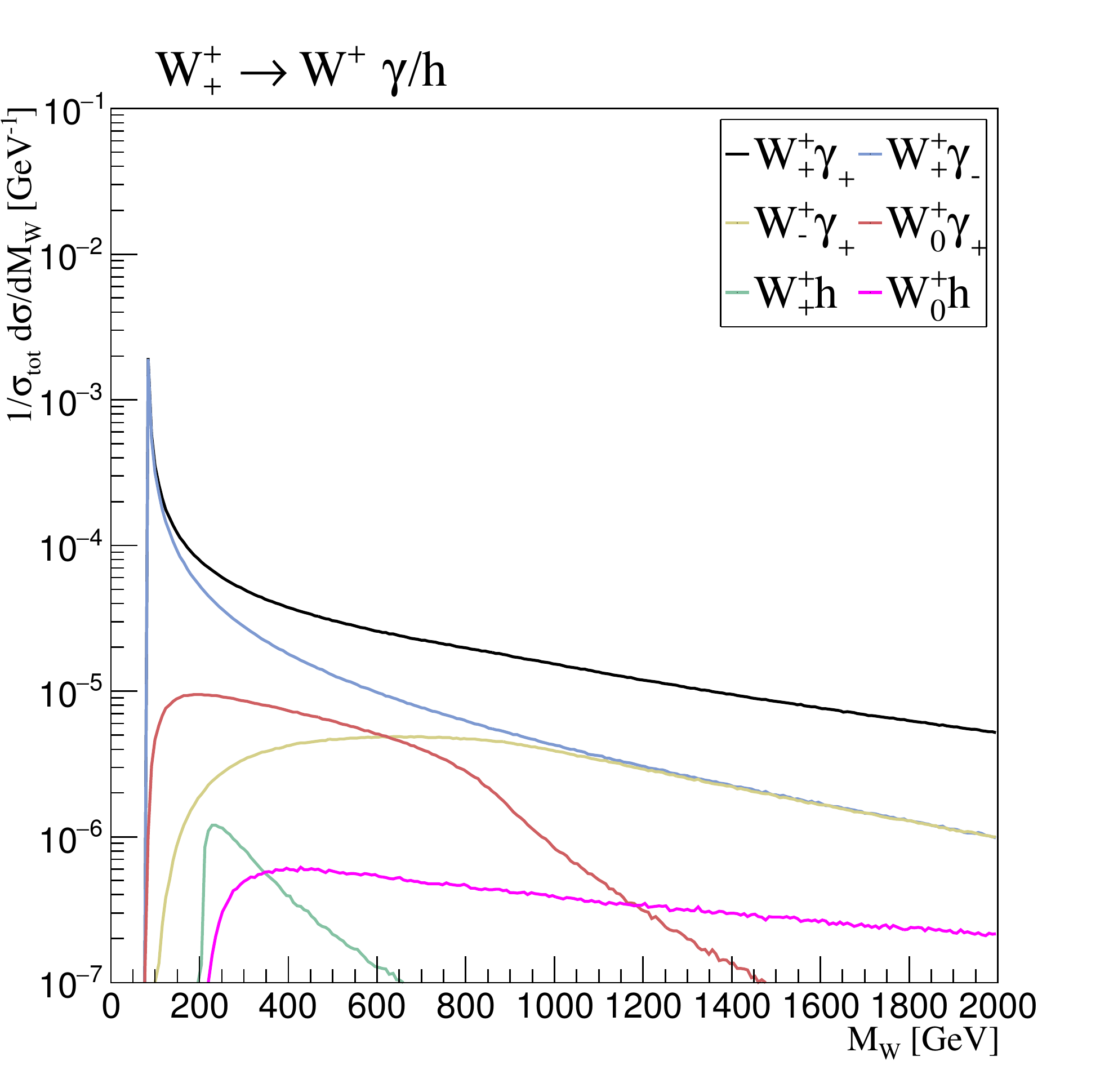}
\caption{Branching spectra of a $1$ TeV $W^+_+$ to fermions (left), $W^+Z$ (right) and $W^+ \gamma/h$ (bottom).}
\label{WEWshowerFigure}
\end{figure*}

\begin{figure*} 
\centering
\begin{minipage}{.5\textwidth}
    \includegraphics[scale=0.4]{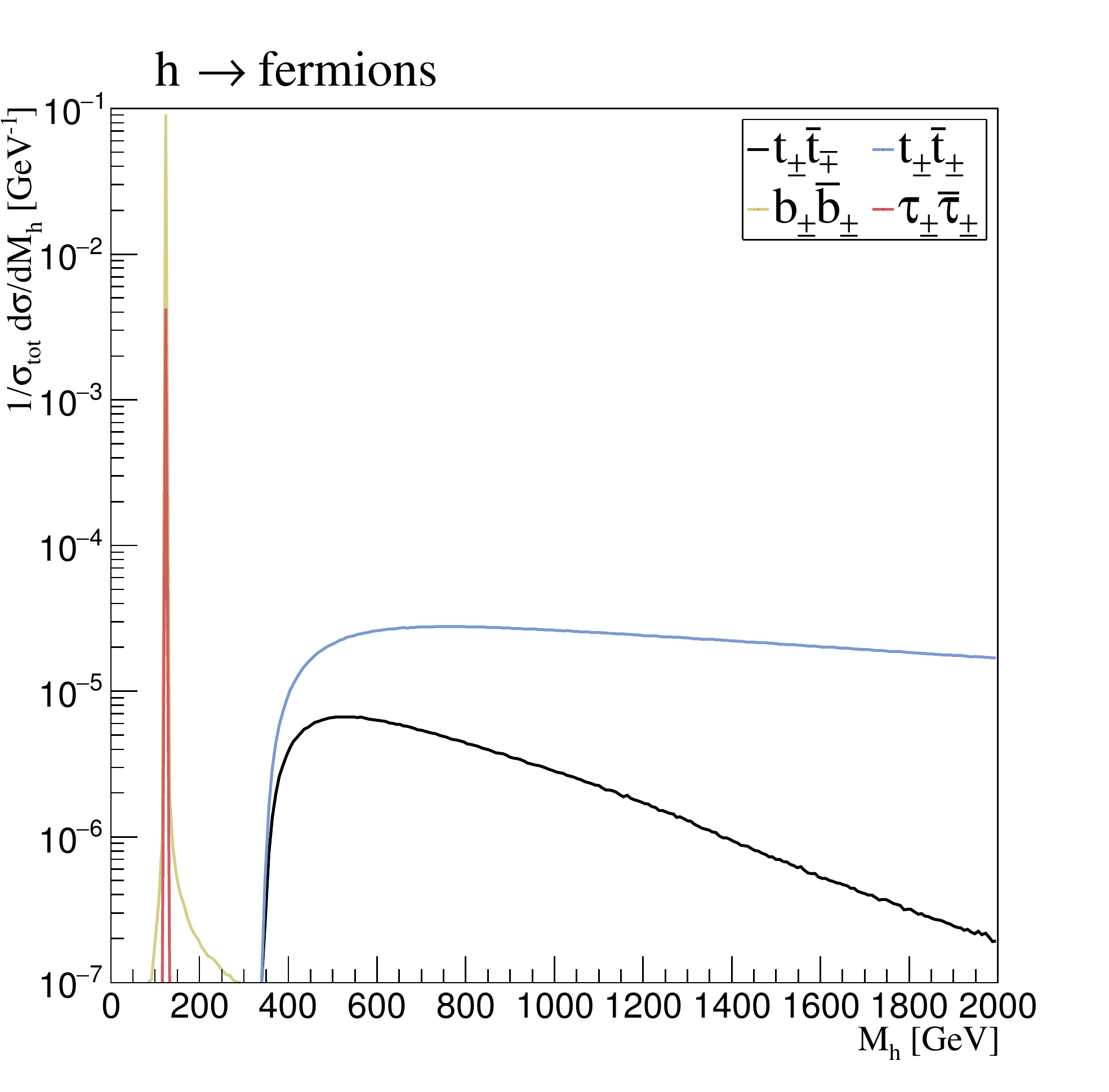}
\end{minipage}%
\begin{minipage}{.5\textwidth} 
    \includegraphics[scale=0.4]{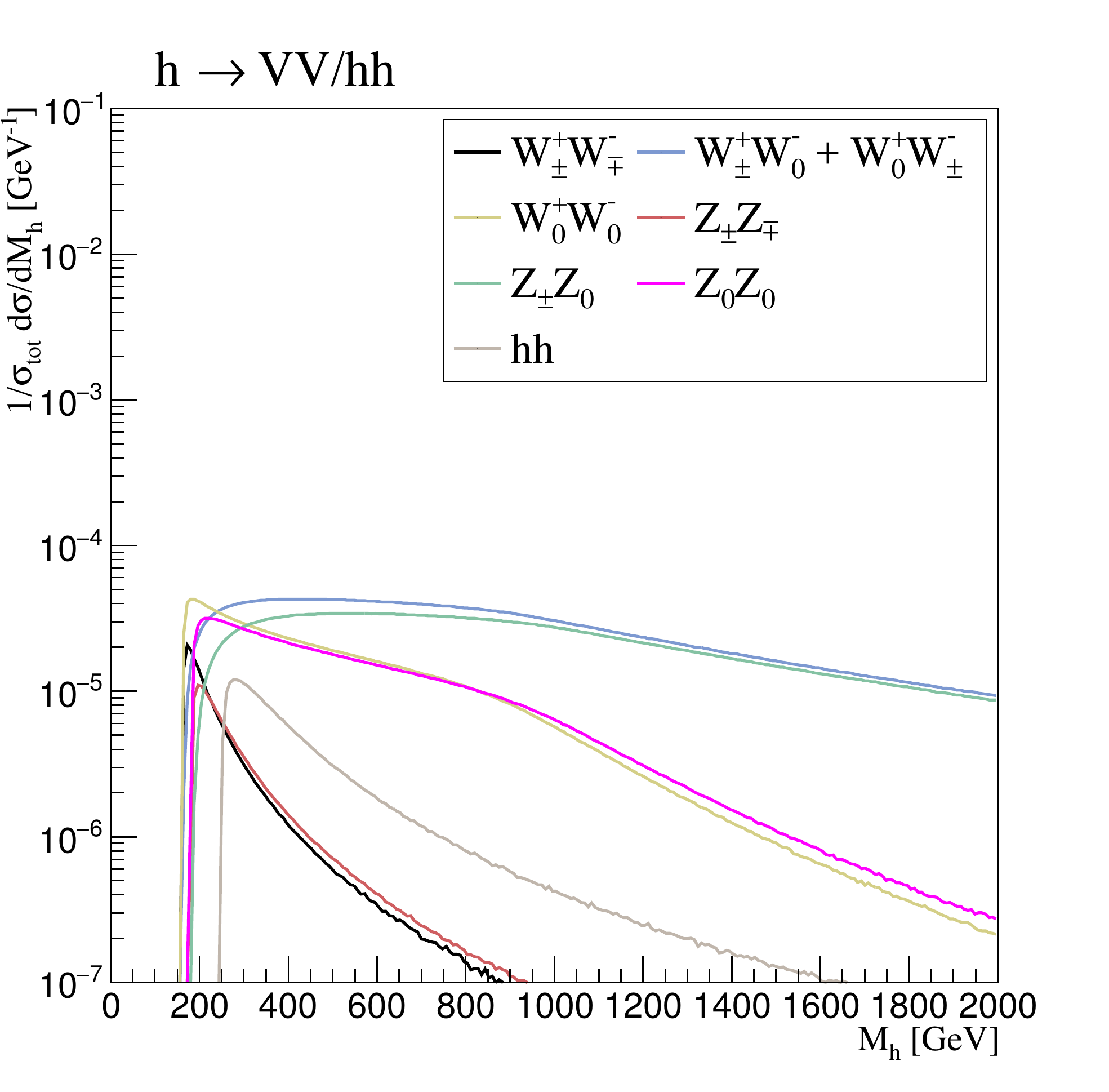}
\end{minipage}
\caption{Branching spectra of a $1$ TeV Higgs boson to fermions (left) and $VV/hh$ (right).}
\label{hEWshowerFigure}
\end{figure*}

In this section, we show electroweak branching spectra of several highly energetic particles as a function of the invariant mass, which closely corresponds with the ordering scale. 
Figures \ref{tauEWshowerFigure}, \ref{topEWshowerFigure}, \ref{WEWshowerFigure} and \ref{hEWshowerFigure} show this for the first branching of a left-handed $\tau$ and top, a transverse $W^+$ boson and a Higgs. 
Note that multiple emission rates are related to these single branching spectra through the usual generalized Poisson statistics associated with shower algorithms due to the multiplicative property of the Sudakov factor \footnote{Although the equivalence is not exact due to recoiler effects and the favour-changing nature of many electroweak branchings.}.
All particles are produced at an energy of $1$ TeV together with a recoiler that is uncharged under electroweak interactions. 
For photon emissions, a cutoff around $\Lambda_{\mbox{\tiny{QCD}}}$ is imposed. 
All other branchings are automatically regulated by the particle masses or by matching to the Breit-Wigner distribution. 

Figure \ref{tauEWshowerFigure} shows the branching spectrum of a negative-helicity $\tau$. 
The two dominant photon production channels are those where the $\tau$ helicity is conserved. 
The mass-suppressed spin-flip mode only contributes at very small invariant masses, as is to be expected from the branching kernel behaviour of $m_{\tau}^2/Q^4$. 
The other spin-flip mode is highly suppressed in the collinear limit as is indicated in Table \ref{Amps1}.
For the emission of other vector bosons, the spin-flip contributions do not become sufficiently enhanced to show up before the kinematic limit is reached. 
The longitudinal vector boson emission channels have a characteristic form which looks very similar for the $W^{-}_{0}$ and the $Z_0$ channels, 
and which becomes comparable to the transverse channels at scales close to the kinematic limit. 

Figure \ref{topEWshowerFigure} shows the branching spectrum of a negative-helicity top. 
The left graph displays the resonance branchings as generated by the sampled matching procedure outlined in subsection \ref{orderingSubsection}.
The right and bottom graphs show all other branchings that are not of the resonance decay type.
Spin-flip modes now show up for $t \rightarrow b W^{+}$, $t \rightarrow t Z$ and $t \rightarrow t \gamma$ due to the large top mass, and they show the expected $m_t^2/Q^4$ scaling with the emission scale. 
The `natural` mode of spin-flip Higgs emission is relatively flat compared with the fermion mass scaling mode of Higgs emission without spin flip.

Figure \ref{WEWshowerFigure} shows the branching spectrum of a transverse $W^{+}$. 
Resonance peaks only appear for decays to negative-helicity states due to their small masses.
The branchings $W^+_+ \rightarrow t \bar{b}$  with a spin-flipped top do occur on the other hand. 
The $W^+_+ \rightarrow W^+ Z$ and $W^+_- \rightarrow W^+ \gamma$ channels are dominated by the all-positive helicity configuration because of its $1/z(1-z)$ scaling in the collinear limit as can be seen in Table \ref{Amps3}.
The modes to opposite transverse helicities are almost identical for the $W^+ Z$ channels due to symmetry in the collinear limit and almost identical mass, 
but they are widely different for low scales in the $W^+ \gamma$ channels. 
This is caused by the $z^3/(1-z)$ and $(1-z)^3/z$ scaling of the collinear limits, where the photon can attain a very small collinear momentum fraction while that of the $W^+$ is constrained by its mass. 
The single-longitudinal channels in $W^+ Z$ are also almost identical for very similar reasons. 
The $W^+_0 Z_0$ is a mode that is related to the Goldstone bosonic part of the $W^+$ and $Z$, and it can be seen to be very similar to the $W_0^+ h$ channel.
On the other hand, the $W^+_+ h$ mode differs significantly from the $W^+_+ Z_0$ channel because it is dominated by the vectorial part of the longitudinal polarization.

Figure \ref{hEWshowerFigure} shows the branching spectrum of a Higgs. 
The only significant resonance decay channels are $b_{\pm} \bar{b}_{\pm}$ and $\tau_{\pm} \bar{\tau}_{\pm}$ as may be expected due to the coupling to the fermion mass and the Higgs spin zero nature.
On the other hand, the mass-suppressed $t_{\pm} \bar{t}_{\mp}$ channel is comparable with the natural $t_{\pm} \bar{t}_{\pm}$ channel.
All channels to $W^+ W^-$ and $ZZ$ are almost identical since their branching kernels only differ in the gauge boson mass and a factor of $1/c_w$ in the coupling. 
Also included is the $h \rightarrow hh$ cubic Higgs coupling which is proportional to the Higgs mass $m_h$, or equivalently the Higgs self-coupling $\lambda$. 
This is the only branching where it makes an appearance, and it can be seen to provide a significant contribution to the total branching rate. 

\subsection{Bosonic Interference}
\begin{figure*} 
\centering
\begin{minipage}{.5\textwidth}
    \includegraphics[scale=0.4]{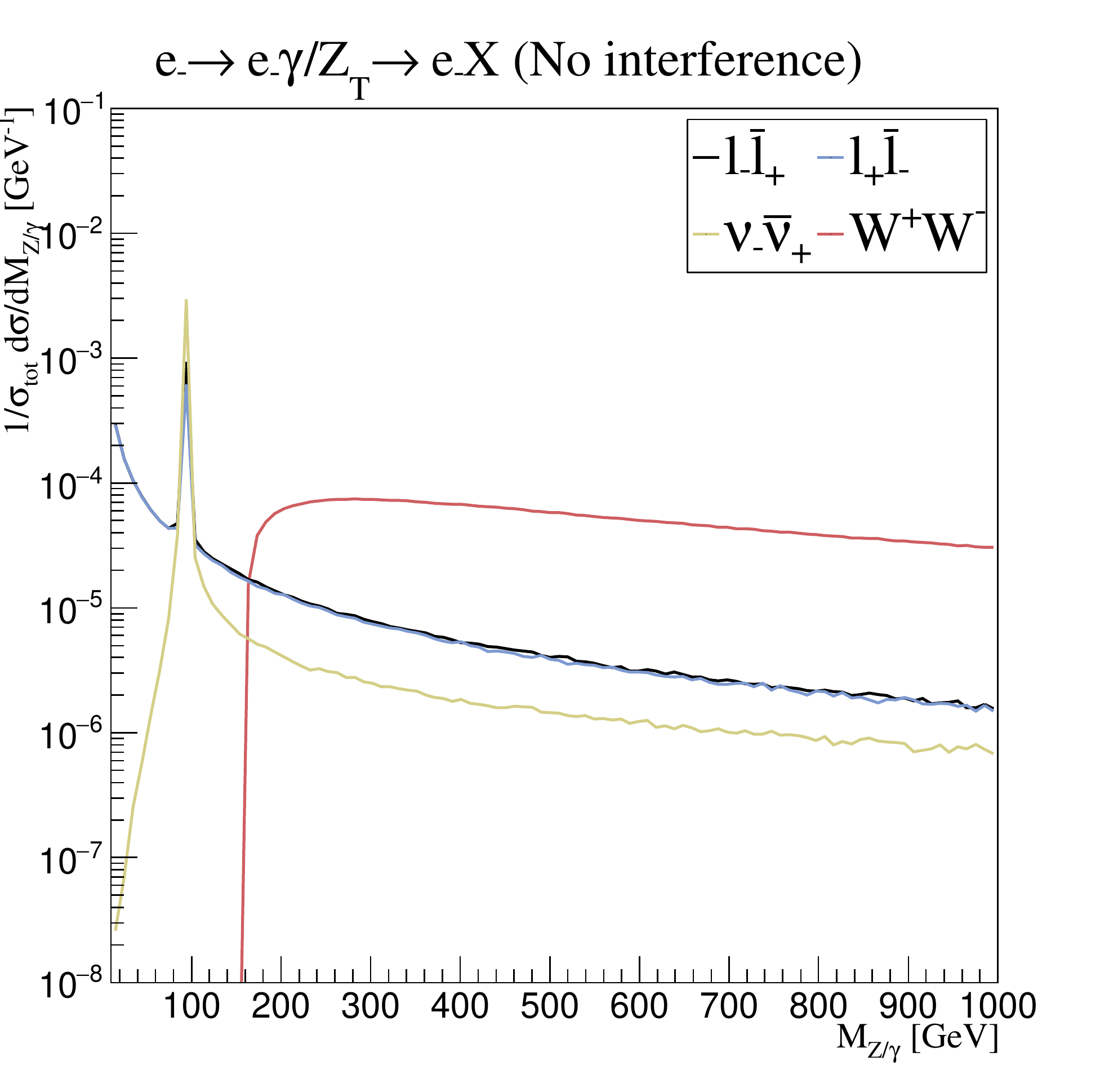}
    \includegraphics[scale=0.4]{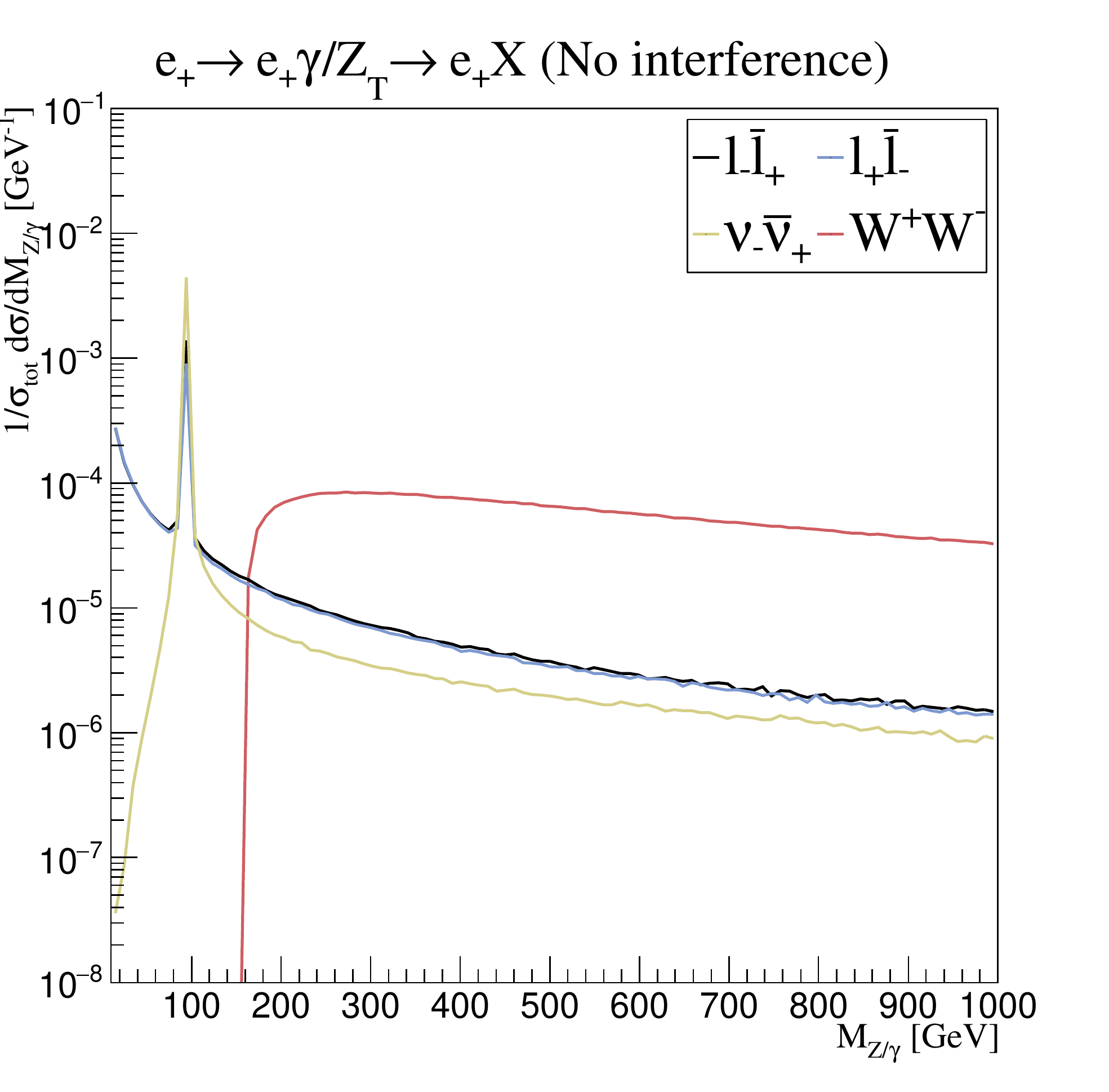}
\end{minipage}%
\begin{minipage}{.5\textwidth} 
    \includegraphics[scale=0.4]{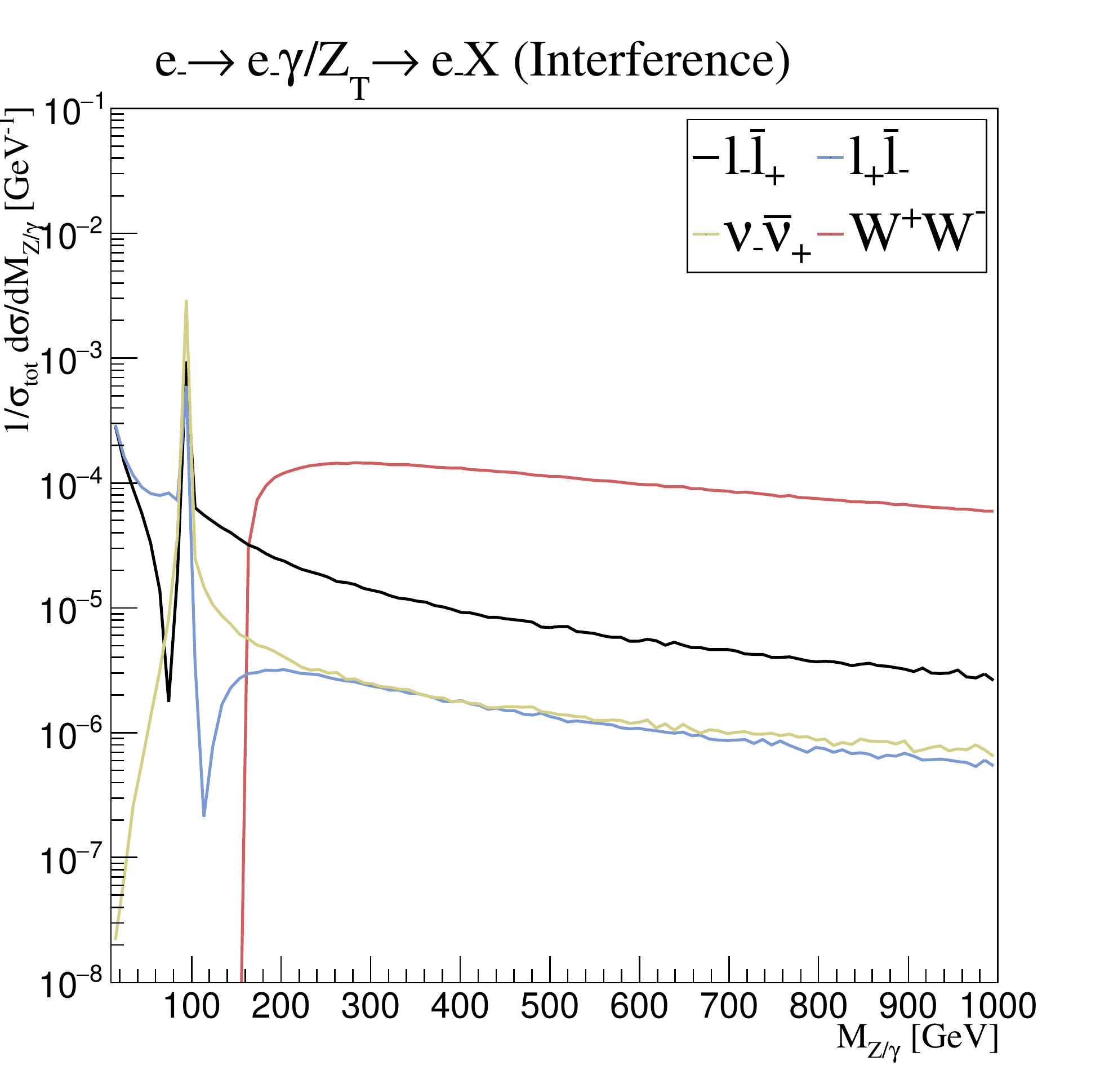}
    \includegraphics[scale=0.4]{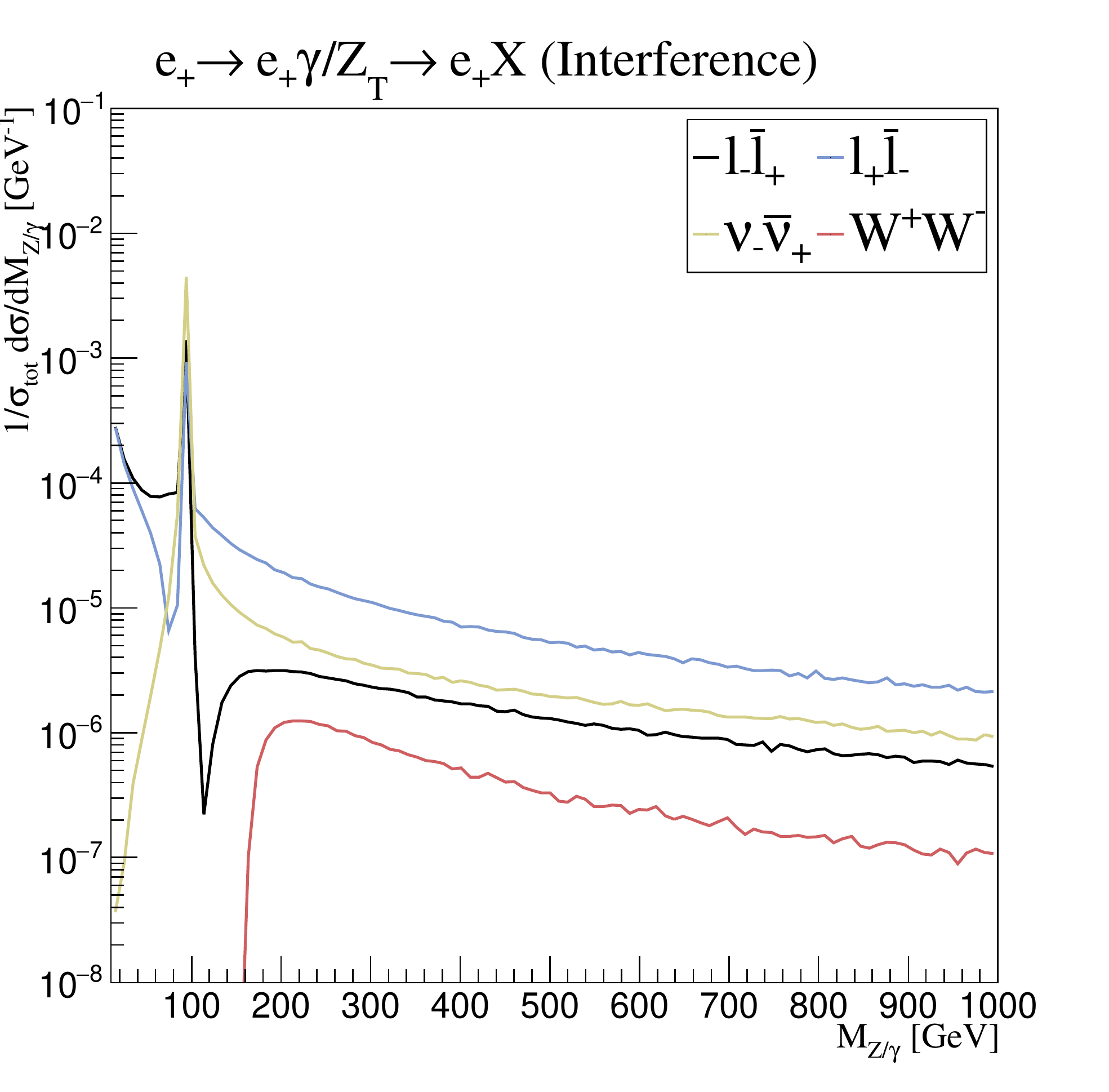}
\end{minipage}
\caption{Differential rates of the shower histories $e_- \rightarrow e_- \gamma / Z_T \rightarrow e_- X$ (top) and $e_+ \rightarrow e_+ \gamma / Z_T \rightarrow e_+ X$ (bottom) 
without (left) and with (right) the bosonic interference correction. 
The showers are initiated from a $10$ TeV electron with a neutral recoiler.}
\label{BosonInterferenceFigure}
\end{figure*}

We now consider the effect of the application of the bosonic interference factor described in section \ref{bosonicInterferenceSubsection}.
Figure \ref{BosonInterferenceFigure} shows rates for the shower histories $e_- \rightarrow e_- \gamma / Z_T \rightarrow e_- X$ and $e_+ \rightarrow e_+ \gamma / Z_T \rightarrow e_+ X$
using a similar setup as in the previous subsection, but starting from a $10$ TeV source electron. 
Multiple interesting features appear when the bosonic interference weight eq.~\eqref{bosonInterferenceCorrectionFactor} is included. 
The most striking difference occurs for the $W^+ W^-$ channel, where the bosonic interference causes an increase in case of the $e_-$, but a major decrease in case of the $e_+$.
This may be understood by considering the structure of the interfering branching amplitudes. 
Factoring out coupling constants and other kinematic components, the interference is proportional to
\begin{equation}
\frac{1}{M_{WW}^2} + \frac{c_w}{s_w} \frac{1}{4 s_w c_w} (1 - 4 s_w^2 - \lambda_e) \frac{1}{M_{WW}^2 - m_z^2 + i m_z \Gamma_z},
\end{equation}
where the factor $c_w/s_w$ comes from the $ZWW$-coupling and $\lambda_e$ is the electron helicity. 
The second term in brackets interferes destructively with the photon contribution for sufficiently large values of $M_{WW}$, 
and the remaining terms in the $Z$ contribution cancel for $\lambda = 1$. 

The effects of the bosonic interference factor on charged fermion rates close to the $Z$ peak may be understood through a similar argument.
The rates close to the $Z$ peak are significantly affected by the simplified and preliminary method of matching to resonance decays as described in section \ref{orderingSubsection}, and will be improved upon in \cite{UpcomingResonance}.

\subsection{Electroweak Corrections to Proton Collision Processes}
We finally consider the parton shower predictions of electroweak corrections to some common proton collision processes at LHC energies and compare with the Pythia electroweak shower \cite{PythiaEW}. 
Since the weak vector bosons produced by the electroweak shower at high energies are massive and thus observable, 
they may provide a rich environment for phenomenological studies including kinematic effects on the hard scattering, 
jet substructure due to vector boson decay inside jet cones and external high-energy jet and lepton production.

\begin{figure*} 
\centering
\begin{minipage}{.5\textwidth}
    \includegraphics[scale=0.4]{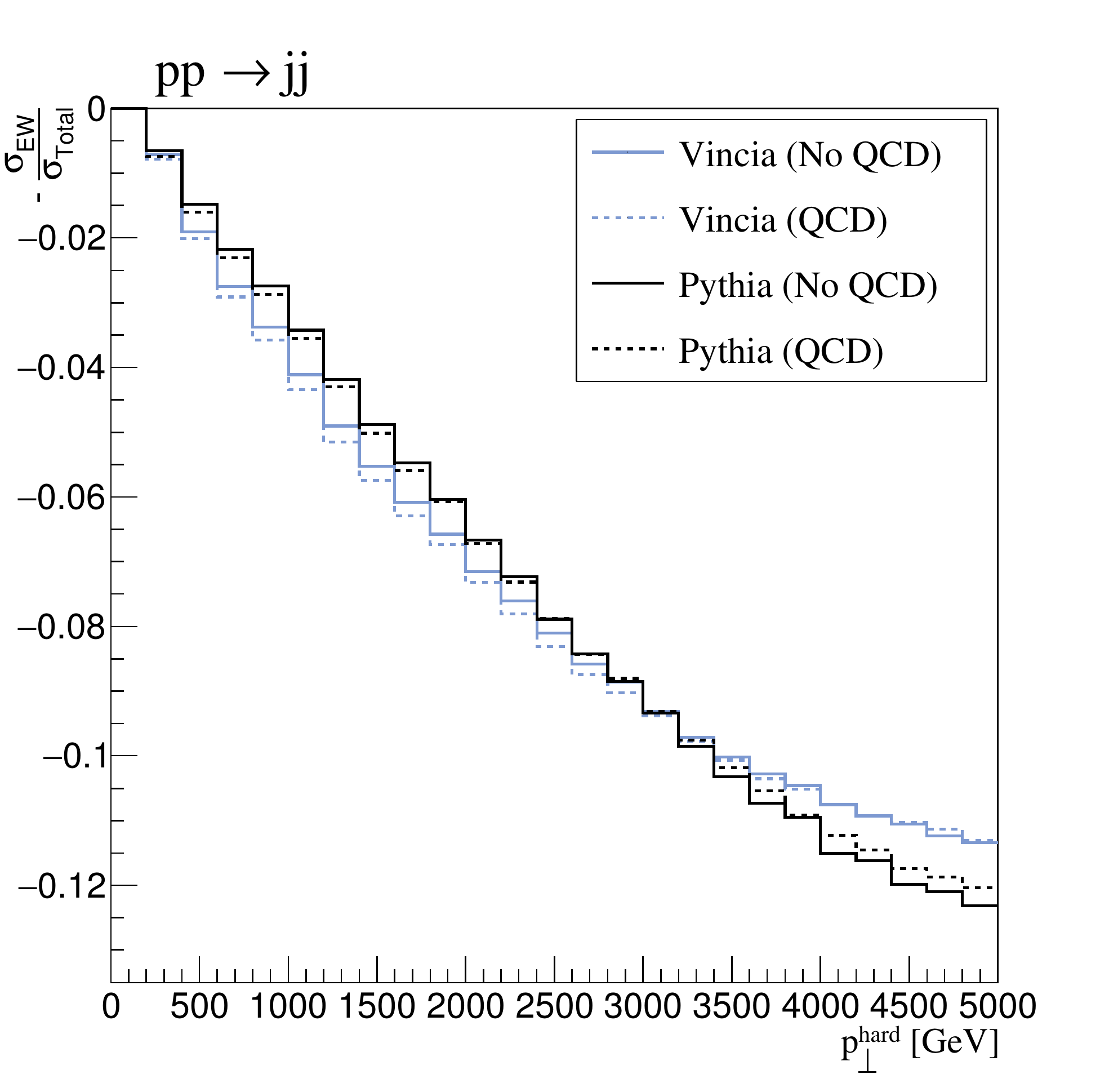}
\end{minipage}%
\begin{minipage}{.5\textwidth} 
    \includegraphics[scale=0.4]{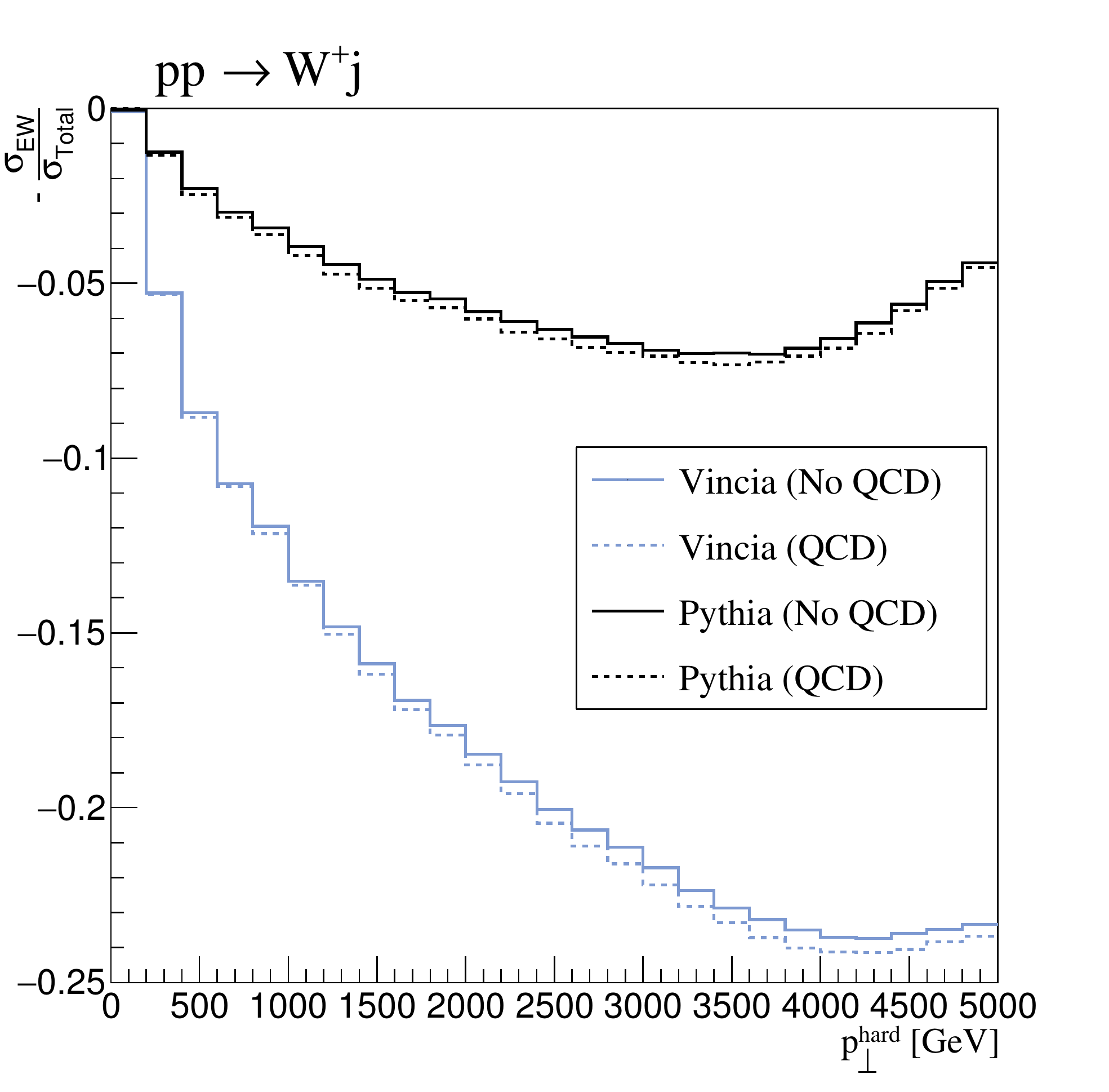}
\end{minipage}
\caption{Electroweak shower approximation of electroweak virtual corrections to exclusive dijet production and exclusive $W^+$ + jet production 
at center-of-mass energy $\sqrt{s} = 14$ TeV as a function of the transverse momentum of the hard scattering process $p_{\perp}^{\mbox{\tiny{Hard}}}$.
The solid line corresponds to the parton shower prediction without the QCD shower, while the dashed line shows the effect of interleaving with QCD radiation.}
\label{Figure14Virtual}
\end{figure*}

With the goal of examining the general significance of electroweak Sudakov effects in common LHC processes, we generate dijet and $W^{+}$ plus jet events 
at $\sqrt{s} = 14$ TeV using the default tune of Pythia 8.2 \cite{Pythia8.2} and the NNPDF2.3 sets \cite{NNPDF2.3}.  
Figure \ref{Figure14Virtual} shows the approximate electroweak virtual corrections as predicted by the Pythia electroweak shower, 
which only incorporates vector boson emission from fermions, and the Vincia electroweak shower as a function of the transverse momentum of the hard scattering. 
The virtual corrections may be estimated by counting the events that contain at least one weak vector boson emission. 
The probability for the shower to produce no additional weak bosons is given by the Sudakov factor 
\begin{equation}
\Delta_{\mbox{\tiny{EW}}} = 1 - \mathcal{O}(\alpha),
\end{equation}
and thus the $\mathcal{O}(\alpha)$ corrections are given by the probability for at least one weak boson emission. 
Virtual corrections to these processes were calculated in for example \cite{diJetLarge1,diJetLarge2} for exclusive dijet production and in \cite{VjLarge2,VjLarge3} for vector boson production. 

For dijet production, the results of the showers are very similar.
In the case of $W^+$ plus jet production the substantial difference between the showers is caused by the absence of the Yang-Mills vector boson coupling in the Pythia shower. 
Furthermore, we find that the contribution to the weak boson emission rates of the initial-state quarks is significantly smaller than that of the final state. 
Because at large $x$ and high scales the PDFs are predominantly quark-like and the hard scattering is dominated by $q \bar{q}' \rightarrow W^+ g$, and the phase space for initial state radiation is small at large $x$, 
virtual corrections decrease at large values of transverse momentum.

Also shown in Figure \ref{Figure14Virtual} are the results of interleaving the electroweak shower with the QCD showers of Pythia and Vincia. 
In the strongly ordered limit shower branchings are unaffected by subsequent branchings, but subleading effects due to the kinematics and the creation of weakly charged quarks still lead to minor differences. 
\begin{figure*} 
\centering
\begin{minipage}{.5\textwidth}
    \includegraphics[scale=0.4]{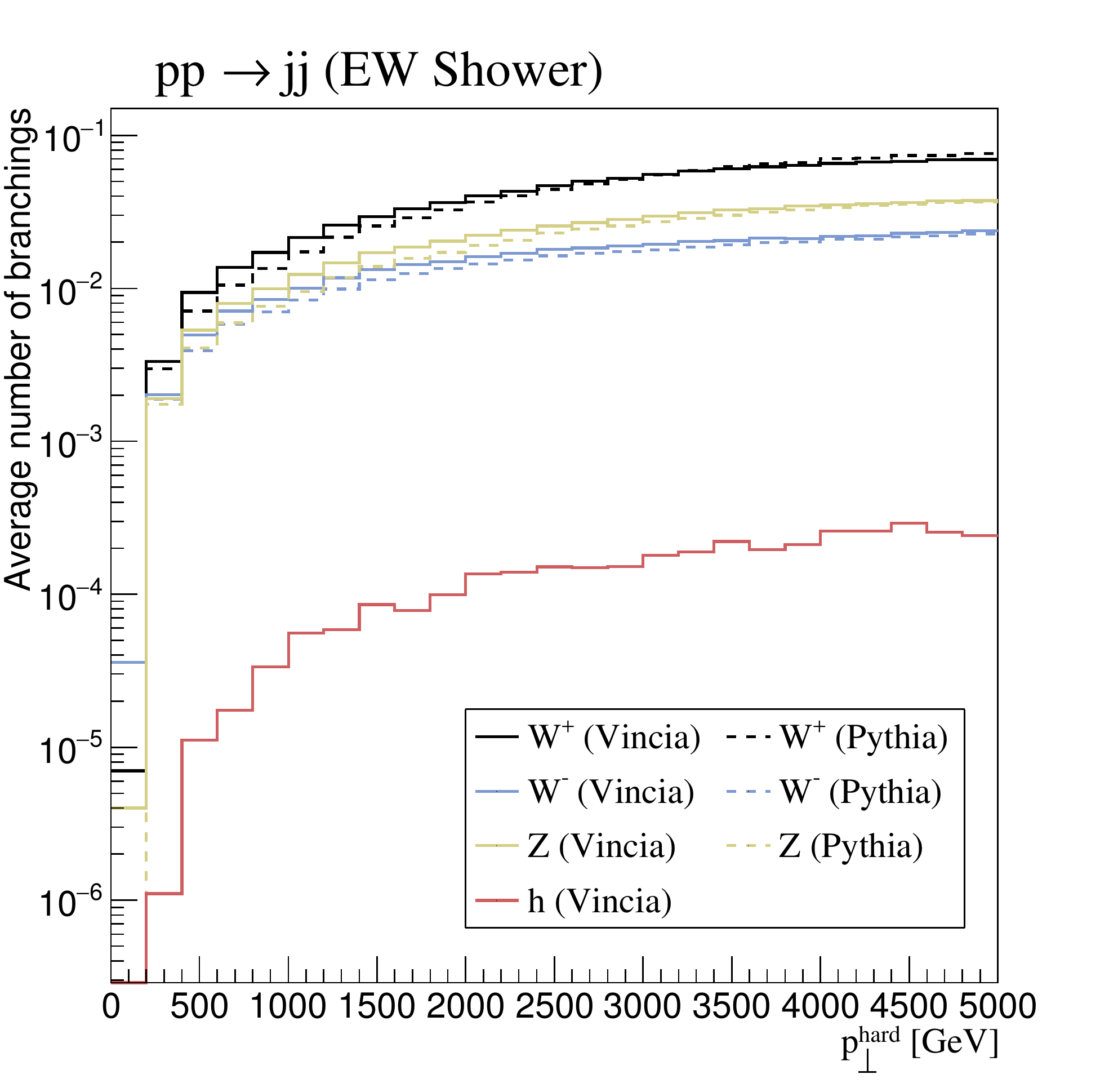}
\end{minipage}%
\begin{minipage}{.5\textwidth} 
    \includegraphics[scale=0.4]{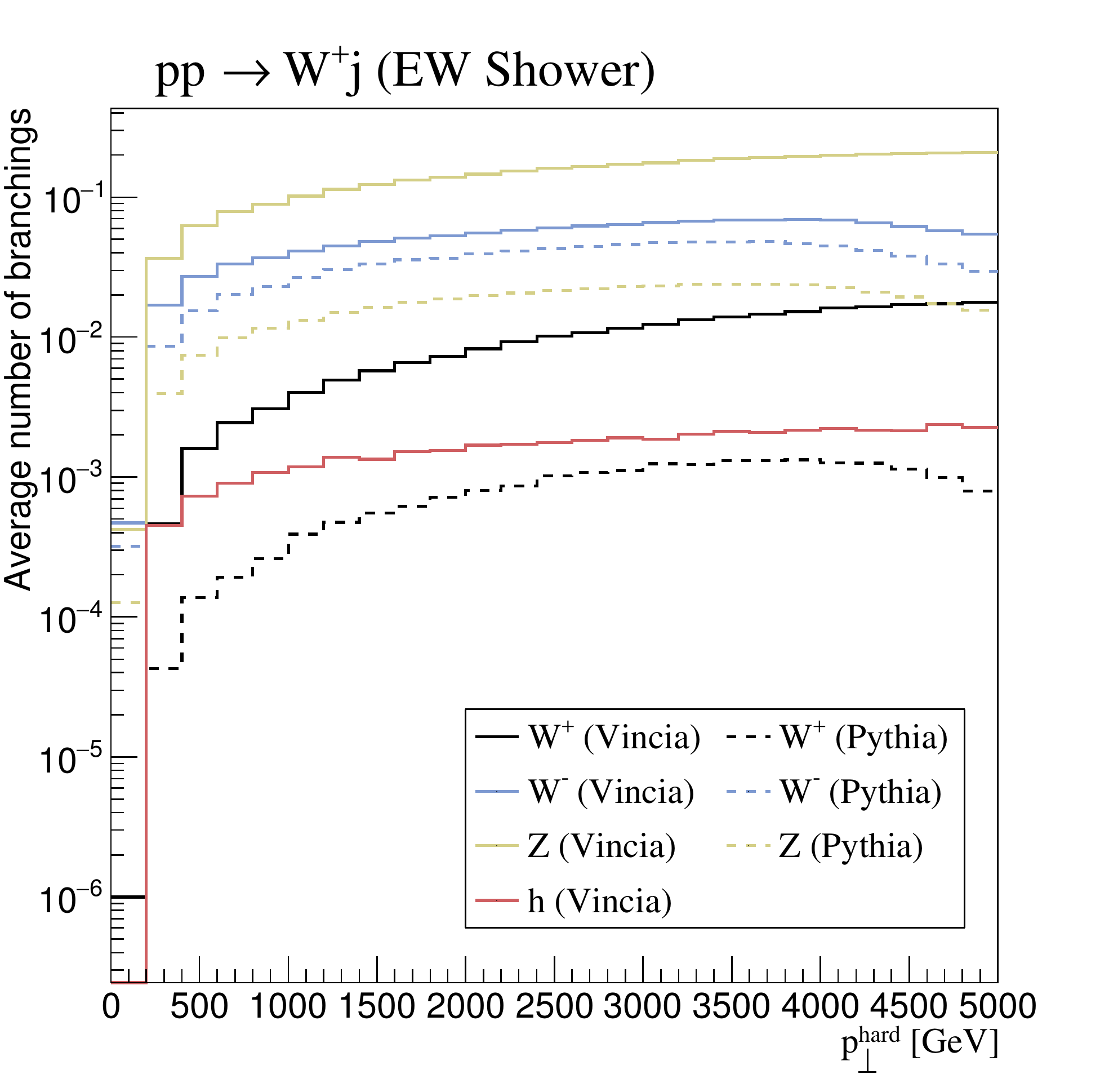}
\end{minipage}
\caption{Average number of weak boson emissions in exclusive dijet production and exclusive $W^+$ + jet production 
at center-of-mass energy $\sqrt{s} = 14$ TeV as a function of the transverse momentum of the hard scattering process $p_{\perp}^{\mbox{\tiny{Hard}}}$.}
\label{Figure14noQCDAverage}
\end{figure*}

Figure \ref{Figure14noQCDAverage} shows the average number of weak boson emissions of the showers. 
In $W^+$ plus jet production the first purely vector boson branching is always $W^+ \rightarrow W^+ Z$ explaining the large increase in the $Z$ boson emission rates.
Similarly, Pythia's $W^{+}$ rate is small since the final-state quark is always down-type. 
The increase in the Vincia shower is thus caused entirely by secondary emissions from prior weak vector boson emissions.

\section{Conclusion and Discussion} \label{conclusion}
The effects of weak corrections in parton shower algorithms are known to become significant already at LHC energies, in particular with the upcoming luminosity upgrade, and will be even more relevant at future colliders.
One of the major challenges of the construction of such a shower is the calculation of the relevant branching kernels, which in this paper was done using the spinor-helicity formalism. 
Compared with QCD, the electroweak theory involves many theoretical subtleties that have to be handled carefully. 
One major issue is the chiral nature of the electroweak theory, which forces the shower to be helicity-dependent and leads to a large number of possible types of branchings. 
In particular, the scalar components of longitudinal polarizations lead to unphysical, unitarity-violating contributions that have to be treated carefully.
The collinear limits of the computed branching kernels are found to be in agreement with the results of \cite{Tweedie}.
The electroweak shower also includes many branchings that would usually be considered to be decays of resonances, in which case the distribution follows a Breit-Wigner peak.
A strategy to match the parton shower to a resonance decay was proposed, but this may likely be improved upon by a better understanding of the interplay between the virtual corrections contained in the Sudakov factor and the decay width. 
A more sophisticated treatment of this matching is beyond the scope of this paper, and will be the topic of future study \cite{UpcomingResonance}.
Further electroweak effects added to the shower include a recoiler selection procedure that compensated for recoiler effects of previous branchings and treatment of bosonic interference effects. 
Results were shown that quantify the general size of electroweak shower corrections at future collider energies and at LHC energies. 

Several features that are currently lacking from the electroweak shower were already pointed out.
They include topics such as soft and spin interference effects, although such issues have similarly not been fully solved in the QCD sector of commonly used shower codes.
However, further issues particular to the electroweak sector still remain. 
One is the inclusion of the CKM quark-mixing matrix \cite{CKM1,CKM2}, which would lead to an even larger number of possible electroweak branchings. 
The impact of the CKM matrix is however not expected to be large since the off-diagonal terms of the third-generation row and column are close to zero. 
The other off-diagonal terms are not as small, but they mix quarks that are treated as massless in Vincia anyway.

One other peculiar property of the electroweak theory is the appearance of Bloch-Nordsieck violation \cite{Bloch-Nordsieck1,EWLarge6}.
The parton shower formalism is fundamentally based on the principle of unitarity and the cancellation of infrared divergences between real and virtual corrections.
Since the electroweak vector bosons are massive, divergences associated with their emission are mass-regulated. 
The flavour-changing nature of $W$-boson emission from the initial state spoils the exact cancellation of the infrared divergences and some mass-regulated logarithms may be left-over. 

There is no straightforward method to incorporate these violations in the shower formalism, since they explicitly break unitarity.
We note that, while Bloch-Nordsieck violations are not particularly significant at the LHC \cite{EWLarge6,diJetLarge2}, they will be important at future collider energies and a comprehensive treatment will be necessary.

Finally, hard processes initiated by vector bosons have been considered for a long time \cite{EffectiveVBA1,EffectiveVBA2}. 
PDF sets with QED corrections have been available for some time \cite{photonPdfs1,photonPdfs2,photonPdfs3,photonPdfs4}, and recent progress was made towards PDFs with complete electroweak corrections \cite{EWPdfs1,EWPdfs2,EWPdfs3}.
The current shower implementation only allows for the emission of vector bosons from the initial state. 
The calculation of the other required initial-state branching kernels is in principle as straightforward as the calculation of those available already, but an implementation in the Pythia framework is likely not simple.

\begin{acknowledgements}
We are grateful to Helen Brooks and Peter Skands for many useful discussions and help with the implementation in the Vincia parton shower.
R. V. acknowledges support by the Foundation for Fundamental Research of Matter (FOM) via program 156 'Higgs as Probe and Portal' and by the Science and Technology Facilities Council (STFC) via grant award ST/P000274/1.
\end{acknowledgements}

\appendix
\section{Relevant Feynman Rules of the Electroweak Theory} \label{EWFeynRules}
\begin{figure}
\begin{align}
&\diag{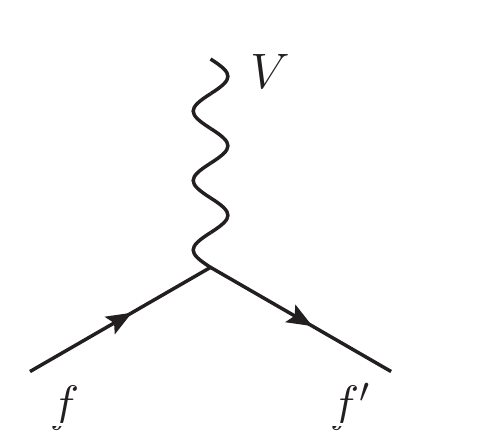}{3}{-1.1} = i (v + a \gamma^5) \gamma^{\mu} \nonumber \\
&\diag{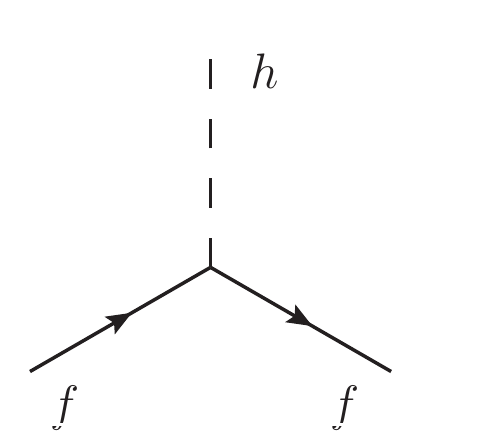}{3}{-1.1} = i \frac{e}{2 s_w} \frac{m_f}{m_W} \nonumber \\
&\diag{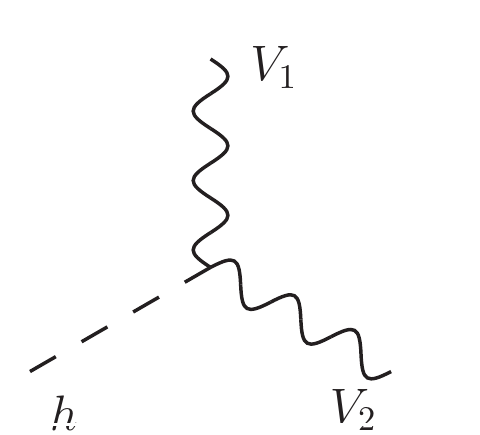}{3}{-1.1} = i g_{h} g^{\mu \nu} \nonumber \\
&\diag{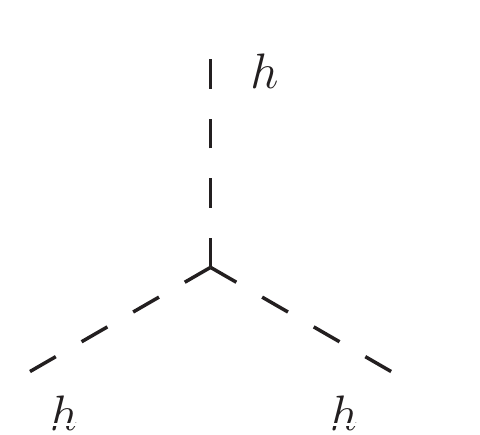}{3}{-1.1} = -i \frac{3}{2} \frac{m_h^2}{m_w s_w}. \nonumber \\
&\diag{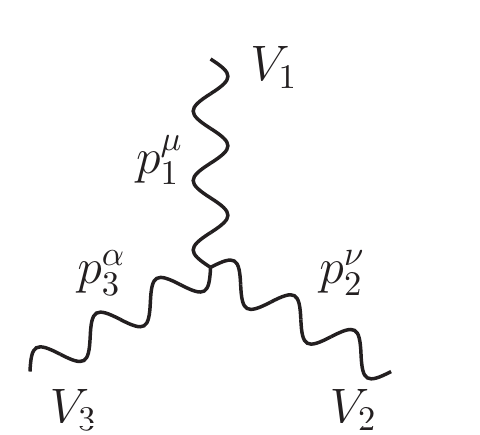}{3.2}{-1.1} = i g_{V} Y(p_1, \mu, p_2, \nu, p_3, \alpha) \nonumber 
\end{align}
\caption{The vertex interactions of the electroweak theory.}
\label{vertexInteractions}
\end{figure}
This appendix lists the Feynman rules of the electroweak theory that are relevant for the calculation of branching amplitudes. 
We elect to make use of a practical notation for the electroweak Feynman rules which makes for simpler results, but obfuscates some of the underlying group structure.  
The relevant vertex interactions are given in Figure \ref{vertexInteractions}, where
\begin{align}
Y(p_1, \mu, p_2, \nu, p_3, \alpha) &= (p_1-p_2)^{\alpha} g^{\mu \nu} + (p_2-p_3)^{\mu} g^{\nu \alpha} \nonumber \\
&+ (p_3 - p_1)^{\nu} g^{\mu \alpha}
\end{align}
is the Yang-Mills vertex.
As usual, the weak mixing angle is defined as 
\begin{equation}
c_w \equiv \cos \theta_w = \frac{m_W}{m_Z} \quad s_w \equiv \sin \theta_w.
\end{equation}
The coupling constants are defined in Table \ref{couplings}.
\begin{table*}
\centering
\scalebox{1.1}{
\begin{tabular}{l|lll|lll|lll}
\multicolumn{1}{l}{} & \multicolumn{3}{c}{$v$} & \multicolumn{3}{c}{$a$} 
      & \multicolumn{2}{c}{\quad} & \multicolumn{1}{c}{$g_V$}\\ \cline{2-7} \cline{10-10}
      & \multicolumn{1}{c}{$\gamma$} & \multicolumn{1}{c}{$W$} & \multicolumn{1}{c}{$Z$} 
      & \multicolumn{1}{|c}{$\gamma$} & \multicolumn{1}{c}{$W$} & \multicolumn{1}{c|}{$Z$}  
      & \multicolumn{1}{c}{\quad} & \multicolumn{1}{c|}{$WW\gamma$} & \multicolumn{1}{c|}{$-e$} \\ \cline{2-7} 
$d$   & \multicolumn{1}{c}{$-\frac{1}{3} e$} & \multicolumn{1}{c}{$\frac{-e}{\sqrt{8} s_w}$} 
      & \multicolumn{1}{c|}{$\frac{-e}{4 s_w c_w} \left(1 - \frac{4}{3} s_w^2 \right)$} & $0$ 
      & \multicolumn{1}{c}{$\frac{-e}{\sqrt{8} s_w}$} & \multicolumn{1}{c|}{$\frac{-e}{4 s_w c_w}$}  
      & \multicolumn{1}{c}{\quad} & \multicolumn{1}{c|}{$WWZ$} & \multicolumn{1}{c|}{$-e \frac{c_w}{s_w}$} \\
$u$   & \multicolumn{1}{c}{$\frac{2}{3} e$} & \multicolumn{1}{c}{$\frac{-e}{\sqrt{8} s_w}$} 
      & \multicolumn{1}{c|}{$\frac{e}{4 s_w c_w} \left(1 - \frac{8}{3} s_w^2 \right)$} & $0$ 
      & \multicolumn{1}{c}{$\frac{-e}{\sqrt{8} s_w}$} & \multicolumn{1}{c|}{$\frac{e}{4 s_w c_w}$}   
      & \multicolumn{1}{c}{\quad} & \multicolumn{1}{c}{\quad} & \multicolumn{1}{c}{$g_h$} \\ \cline{10-10}
$e$   & \multicolumn{1}{c}{$-e$} & \multicolumn{1}{c}{$\frac{-e}{\sqrt{8} s_w}$} 
      & \multicolumn{1}{c|}{$\frac{-e}{4 s_w c_w} \left(1 - 4 s_w^2\right)$} & $0$ 
      & \multicolumn{1}{c}{$\frac{-e}{\sqrt{8} s_w}$} & \multicolumn{1}{c|}{$\frac{-e}{4 s_w c_w}$}   
      & \multicolumn{1}{c}{\quad} & \multicolumn{1}{c|}{$hWW$} & \multicolumn{1}{c|}{$-e\frac{m_w}{s_w}$} \\
$\nu$ & \multicolumn{1}{c}{$0$} & \multicolumn{1}{c}{$\frac{-e}{\sqrt{8} s_w}$} 
      & \multicolumn{1}{c|}{$\frac{e}{4 s_w c_w}$} & $0$ 
      & \multicolumn{1}{c}{$\frac{-e}{\sqrt{8} s_w}$} & \multicolumn{1}{c|}{$\frac{e}{4 s_w c_w}$}    
      & \multicolumn{1}{c}{\quad} & \multicolumn{1}{c|}{$hZZ$} & \multicolumn{1}{c|}{$-e\frac{m_w}{s_w}$}
\end{tabular}}
\captionsetup{justification=centering}
\caption{Values of the coupling constants}
\label{couplings}
\end{table*}

\section{Branching amplitudes} \label{EWbranchingAmplitudes}
\allowdisplaybreaks
All branching amplitudes are multiplied by a propagator factor $1/Q^2$ where 
\begin{align}
Q^2 = 
\begin{cases}
2 p_i{\cdot}p_j + m_i^2 + m_j^2 - m_{ij}^2  & \text{(Final State)} \\
2 p_a{\cdot}p_j - m_j^2                     & \text{(Initial State)}.
\end{cases}
\end{align}
\subsection{Vector Boson Emission}
We define the prefactors 
\begin{align}
A^{\mbox{\tiny{emit}}}_{\perp} &= \frac{1}{2 \sqrt{2}} \frac{\lambda}{\sqrt{p_{i}{\cdot}k_{i}} \sqrt{p_{ij}{\cdot}k_{ij}} \, p_j{\cdot}k_j} \nonumber \\
A^{\mbox{\tiny{emit}}}_{L} &= \frac{1}{2} \frac{1}{m_j} \frac{1}{\sqrt{p_i {\cdot} k_i} \sqrt{p_{ij}{\cdot}k_{ij}}}.
\end{align}
\subsubsection{Vector Boson Emission from Fermion}
\begin{align}
M&^{f \rightarrow f'V}(\lambda, \lambda, \lambda) = A^{\mbox{\tiny{emit}}}_{\perp} \nonumber \\
\times & 
\begin{aligned}[t]
    &\bigg[(v-\lambda a) S_{-\lambda}(k_i, p_i, p_j, k_j)S_{-\lambda} (k_j, p_{ij}, k_{ij}) \nonumber \\
    &+ (v+\lambda a) m_i m_{ij} S_{-\lambda}(k_i, k_j) S_{-\lambda}(k_j, p_j, k_{ij}) \bigg]
\end{aligned} \nonumber \\
M&^{f \rightarrow f'V}(\lambda, \lambda, -\lambda) = A^{\mbox{\tiny{emit}}}_{\perp} \nonumber \\
\times & 
\begin{aligned}[t]
    & \bigg[(v-\lambda a) S_{-\lambda}(k_i, p_i, k_j) S_{-\lambda}(k_j, p_j, p_{ij}, k_{ij}) \nonumber \\
    & + (v+\lambda a) m_i m_{ij} S_{-\lambda}(k_i, p_j, k_j) S_{-\lambda}(k_j, k_{ij}) \bigg] \nonumber \\
\end{aligned} \nonumber \\
M&^{f \rightarrow f'V}(\lambda, -\lambda, \lambda) = A^{\mbox{\tiny{emit}}}_{\perp} \nonumber \\
\times &
\begin{aligned}[t]
    &\bigg[m_{ij} (v+\lambda a) S_{\lambda}(k_i, p_i, k_j) S_{-\lambda}(k_j, p_j, k_{ij}) \nonumber \\
    &- m_i (v-\lambda a) S_{\lambda}(k_i, p_j, k_j) S_{-\lambda}(k_j, p_{ij}, k_{ij}) \bigg]
\end{aligned} \nonumber \\
M&^{f \rightarrow f'V}(\lambda, -\lambda, -\lambda) = A^{\mbox{\tiny{emit}}}_{\perp} \nonumber \\
\times &
\begin{aligned}[t]
&\bigg[m_{ij} (v+\lambda a) S_{\lambda}(k_i, p_i, p_j, k_j) S_{\lambda}(k_j, k_{ij}) \nonumber \\
&- m_i (v-\lambda a) S_{\lambda}(k_i, k_j) S_{\lambda}(k_j, p_j, p_{ij}, k_{ij}) \bigg]
\end{aligned} \nonumber \\
M&^{f \rightarrow f'V}(\lambda, \lambda, 0) = A^{\mbox{\tiny{emit}}}_{L}  \nonumber \\
\times &
\begin{aligned}[t]
&\bigg[ S_{-\lambda}(k_i, (v-\lambda a)(m_{ij}^2 p_i - m_i^2 p_{ij}) \nonumber \\
&+ (v+\lambda a)m_i m_{ij} p_j, k_{ij}) \nonumber \\
&- \frac{m_j^2}{p_j{\cdot}k_j}  \bigg( (v-\lambda a) S_{-\lambda}(k_i, p_i, k_j, p_{ij}, k_{ij}) \nonumber \\
&+ (v + \lambda a) m_{ij} m_i S_{-\lambda}(k_i, k_j, k_{ij}) \bigg) \bigg]
\end{aligned} \nonumber \\
M&^{f \rightarrow f'V}(\lambda, -\lambda, 0) = A^{\mbox{\tiny{emit}}}_{L} \nonumber \\
\times &
\begin{aligned}[t]
&\bigg[ m_i (v-\lambda a)  S_{-\lambda}(k_i, p_j - \frac{m_j^2}{p_j{\cdot}k_j} k_j, p_{ij}, k_{ij}) \nonumber \\
&+ m_{ij} (v+\lambda a) S_{-\lambda} ( k_i, p_i, p_j - \frac{m_j^2}{p_j{\cdot}k_j} k_j, k_{ij}) \bigg]
\end{aligned}
\end{align}

\subsubsection{Vector Boson Emission from Antifermion}
\begin{align}
M&^{\bar{f} \rightarrow \bar{f}'V}(\lambda, \lambda, \lambda) = A^{\mbox{\tiny{emit}}}_{\perp} \nonumber \\
\times &
\begin{aligned}[t]
& \bigg[(v + \lambda a) S_{\lambda}(k_{ij}, p_{ij}, k_j) S_{-\lambda}(k_j, p_j, p_i, k_i) \nonumber \\
& + (v - \lambda a) m_i m_{ij} S_{\lambda}(k_{ij}, p_j, k_j) S_{-\lambda} (k_j, k_i) \bigg]
\end{aligned} \nonumber \\
M&^{\bar{f} \rightarrow \bar{f}'V}(\lambda, \lambda, -\lambda) = A^{\mbox{\tiny{emit}}}_{\perp} \nonumber \\
\times &
\begin{aligned}[t]
&\bigg[(v + \lambda a) S_{\lambda}(k_{ij}, p_j, k_j) S_{-\lambda}(k_j, p_i, k_i) \nonumber \\
& + (v - \lambda a) m_i m_{ij} S_{\lambda}(k_{ij}, k_j) S_{-\lambda}(k_j, p_i, k_i) \bigg]
\end{aligned} \nonumber \\
M&^{\bar{f} \rightarrow \bar{f}'V}(\lambda, -\lambda, \lambda) = A^{\mbox{\tiny{emit}}}_{\perp} \nonumber \\
\times &
\begin{aligned}[t]
&\bigg[m_{ij} (v - \lambda a) S_{\lambda}(k_{ij}, p_j, k_j) S_{-\lambda}(k_j, p_i, k_i) \nonumber \\
& - m_i (v + \lambda a) S_{\lambda}(k_{ij}, p_{ij}, k_j) S_{-\lambda}(k_j, p_j, k_i) \bigg]
\end{aligned} \nonumber \\
M&^{\bar{f} \rightarrow \bar{f}'V}(\lambda, -\lambda, -\lambda) = A^{\mbox{\tiny{emit}}}_{\perp} \nonumber \\
\times &
\begin{aligned}[t]
&\bigg[ m_{ij} (v - \lambda a) S_{\lambda}(k_{ij}, k_j) S_{-\lambda}(k_j, p_j, p_i, k_i) \nonumber \\
& - m_i (v + \lambda a) S_{\lambda}(k_{ij}, p_{ij}, p_j, k_j) S_{-\lambda}(k_j, k_i) \bigg]
\end{aligned} \nonumber \\
M&^{\bar{f} \rightarrow \bar{f}'V}(\lambda, \lambda, 0) = A^{\mbox{\tiny{emit}}}_{L} \nonumber \\
\times &
\begin{aligned}[t]
&\bigg[ S_{\lambda}(k_{ij}, (v + \lambda a)( m_{ij}^2 p_i - m_i^2 p_{ij}) \nonumber \\
& + (v - \lambda a) m_{ij} m_i p_j, k_i) \nonumber \\
& - \frac{m_j^2}{p_j{\cdot}k_j} \bigg( (v + \lambda a) S_{\lambda}(k_{ij}, p_{ij}, k_j, p_i, k_i) \nonumber \\
& + ( v - \lambda a) m_{ij} m_i S_{\lambda}(k_{ij}, k_j, k_i) \bigg) \bigg]
\end{aligned} \nonumber \\
M&^{\bar{f} \rightarrow \bar{f}'V}(\lambda, -\lambda, 0) = A^{\mbox{\tiny{emit}}}_{L} \nonumber \\
\times &
\begin{aligned}[t]
& \bigg[ m_i (v + \lambda a) S_{\lambda}(k_{ij}, p_{ij}, p_j - \frac{m_j^2}{p_j{\cdot}k_j} k_j, k_i) \nonumber \\
& + m_{ij} (v - \lambda a) S_{\lambda}(k_{ij}, p_j - \frac{m_j^2}{p_j{\cdot}k_j} k_j, p_i, k_i) \bigg]
\end{aligned} \nonumber \\
\end{align}

\subsubsection{Vector Boson Emission from Vector Boson} \label{appendixWithPolarizations}
The branching amplitude can be written as
\begin{align} 
M&^{\mbox{\tiny{V$\rightarrow$V'V''}}}(\lambda_{ij}, \lambda_i, \lambda_j) \nonumber \\ 
&= - 2 g_{V} \left( p_j{\cdot}\epsilon_i \, \epsilon_j {\cdot} \bar{\epsilon}_{ij} - p_i {\cdot} \epsilon_{j} \, \epsilon_i {\cdot} \bar{\epsilon}_{ij} + p_i {\cdot} \bar{\epsilon}_{ij} \, \epsilon_i {\cdot} \epsilon_j \right)
\end{align}
To compute the branching amplitude for all helicity configurations, we write out all possible products of momenta and polarization vectors
\begin{align} 
\epsilon&_{\lambda}(p_a) {\cdot} \epsilon_{\lambda}(p_b) \nonumber \\
&= - \frac{1}{4} \frac{1}{p_a {\cdot} k_a \, p_b {\cdot k_b}} S_{-\lambda}(k_a, p_a, p_b, k_b) S_{\lambda}(k_b, k_a) \nonumber \\
\epsilon&_{\lambda}(p_a) {\cdot} \epsilon_{-\lambda}(p_b) \nonumber \\
&= - \frac{1}{4} \frac{1}{p_a {\cdot} k_a \, p_b {\cdot k_b}} S_{\lambda}(k_a, p_a, k_b) S_{-\lambda}(k_a, p_b, k_b) \nonumber \\
\epsilon&_{\lambda}(p_a) {\cdot} \epsilon_0 (p_b) \nonumber \\
&= \frac{\lambda}{2\sqrt{2}} \frac{1}{m_b} \frac{1}{p_a {\cdot} k_a} 
\begin{aligned}[t]
&\bigg(S_{-\lambda}(k_a, p_a, p_b, k_a) \nonumber \\ 
&- \frac{m_b}{p_b {\cdot} k_b} S_{-\lambda}(k_a, p_a, k_b, k_a) \bigg) \nonumber
\end{aligned} \nonumber \\
\epsilon&_0 (p_a) {\cdot} \epsilon_0 (p_b) \nonumber \\
&= \frac{1}{m_a m_b} 
\begin{aligned}[t]
&\bigg( p_a {\cdot} p_b - \frac{m_a}{p_a {\cdot} k_a} k_a \nonumber \\
&- \frac{m_b}{p_b {\cdot} k_b} k_b + \frac{m_a}{p_a {\cdot} k_a} \frac{m_b}{p_b {\cdot} k_b} k_a {\cdot} k_b \bigg) \nonumber \\
\end{aligned} \nonumber \\
\epsilon&_\lambda (p_a) {\cdot} p_b \nonumber \\
&= \frac{\lambda}{\sqrt{2}} \frac{1}{p_a {\cdot} k_a} S_{-\lambda}(k_a, p_a, p_b, k_a) \nonumber \\
\epsilon&_0 (p_a) {\cdot} p_b \nonumber \\
&= \frac{1}{m_a} \left(p_a {\cdot} p_b - \frac{m_a^2}{p_a {\cdot} k_a}\right)
\end{align}
The unitarity-violating terms are then removed by the substitutions
\begin{align}
2 p_i {\cdot} p_j &\rightarrow m_{ij}^2 - m_i^2 - m_j^2 \nonumber \\
2 p_{ij} {\cdot} p_i &\rightarrow m_{ij}^2 + m_i^2 - m_j^2 \nonumber \\
2 p_{ij} {\cdot} p_j &\rightarrow m_{ij}^2 - m_i^2 + m_j^2
\end{align}
The same substitutions are used in the computation of amplitudes involving a Higgs and two vector bosons, where similar products of polarization vectors occur.
\subsection{Higgs Emission}
\subsubsection{Higgs Emission from Fermion}
\begin{align}
M&^{f \rightarrow f h}(\lambda, -\lambda, h) = \frac{e}{4 s_w} \frac{m_i}{m_w} \frac{1}{\sqrt{p_{ij}{\cdot}k_{ij}} \sqrt{p_{i}{\cdot}k_i}} \nonumber \\
&\times \bigg[ S_{-\lambda}(k_i, p_i, p_{ij}, k_{ij}) + m_i^2 S_{-\lambda}(k_i, k_{ij}) \bigg] \nonumber \\
M&^{f \rightarrow f h}(\lambda, \lambda, h) = \frac{e}{4 s_w} \frac{m_i^2}{m_w} \frac{1}{\sqrt{p_{ij}{\cdot}k_{ij}} \sqrt{p_{i}{\cdot}k_i}} \nonumber \\
&\times S_{-\lambda}(k_i, p_i + p_{ij}, k_{ij})
\end{align}

\subsubsection{Higgs Emission from Antifermion}
\begin{align}
M&^{\bar{f} \rightarrow \bar{f} h}(\lambda, -\lambda, h) = \frac{e}{4 s_w} \frac{m_i}{m_w} \frac{1}{\sqrt{p_{ij}{\cdot}k_{ij}} \sqrt{p_{i}{\cdot}k_i}} \nonumber \\
&\times \bigg[ S_{\lambda}(k_{ij}, p_{ij}, p_i, k_i) + m_i^2 S_{\lambda}(k_{ij}, k_i) \bigg] \nonumber \\
M&^{\bar{f} \rightarrow \bar{f} h}(\lambda, \lambda, h) = \frac{e}{4 s_w} \frac{m_i^2}{m_w} \frac{1}{\sqrt{p_{ij}{\cdot}k_{ij}} \sqrt{p_{i}{\cdot}k_i}} \nonumber \\
&\times S_{\lambda}(k_{ij}, p_i + p_{ij}, k_i)
\end{align}

\subsubsection{Higgs Emission from Vector Boson}
\begin{align}
M&^{V \rightarrow V h}(\lambda, \lambda, h) = - \frac{g_{h}}{4} \frac{1}{p_{ij}{\cdot}k_{ij} \, p_i{\cdot}k_i} \nonumber \\
&\times S_{-\lambda}(k_{ij}, p_{ij}, k_i) S_{-\lambda}(k_{ij}, p_i, k_i) \nonumber \\
M&^{V \rightarrow V h}(\lambda, -\lambda, h)= - \frac{g_{h}}{4} \frac{1}{p_{ij}{\cdot}k_{ij} \, p_i{\cdot}k_i} \nonumber \\
&\times S_{-\lambda}(k_i, k_{ij}) S_{-\lambda}(k_{ij}, p_{ij}, p_i, k_i) \nonumber \\
M&^{V \rightarrow V h}(0, \lambda, h) = - \frac{g_{h}}{2\sqrt{2}} \frac{1}{m_{ij}} \frac{\lambda}{p_i{\cdot}k_i} \nonumber \\
&\times S_{-\lambda}(k_i, p_i, p_{ij} - \frac{m_{ij}^2}{p_{ij}{\cdot}k_{ij}} k_{ij}, k_i) \nonumber \\
M&^{V \rightarrow V h}(\lambda, 0, h) = - \frac{g_{h}}{2\sqrt{2}} \frac{1}{m_i} \frac{\lambda}{p_{ij}{\cdot}k_{ij}} \nonumber \\
&\times S_{-\lambda}(k_{ij}, p_{ij}, p_i  - \frac{m_i^2}{p_i{\cdot}k_i} k_i, k_{ij}) \nonumber \\
M&^{V \rightarrow V h}(0, 0, h) = - \frac{g_{h}}{m_{ij}^2}  \nonumber \\
&\times \bigg[\frac{1}{2}m_j^2 + m_{ij}^2 \left(\frac{p_i{\cdot}k_i}{p_{ij}{\cdot}k_{ij}} + \frac{p_j{\cdot}k_j}{p_i{\cdot}k_i} \right)\bigg]
\end{align}

\subsubsection{Higgs Emission from Higgs}
\begin{equation}
M^{h \rightarrow h h}(h, h, h) = \frac{3}{2} \frac{m_{ij}^2}{m_w s_w}
\end{equation}

\subsection{Vector Boson Splitting}
\subsubsection{Vector Boson Splitting to Fermion-antifermion}
Defining the prefactors 
\begin{align}
A^{\mbox{\tiny{split}}}_{\perp} &= -\frac{1}{2 \sqrt{2}} \frac{\lambda}{p_{ij}{\cdot}k_{ij} \sqrt{p_i{\cdot}k_i} \sqrt{p_j{\cdot}k_j}} \nonumber \\
A^{\mbox{\tiny{split}}}_{L} &= \frac{1}{2} \frac{1}{m_{I}} \frac{1}{\sqrt{p_i{\cdot}k_i} \sqrt{p_j{\cdot}k_j}}
\end{align}
the branching amplitudes are 
\begin{align}
M&^{V \rightarrow f\bar{f}}(\lambda, \lambda, -\lambda) = A^{\mbox{\tiny{split}}}_{\perp} \nonumber \\
\times &
\begin{aligned}[t]
& \bigg[ (v - \lambda a) S_{-\lambda}(k_i, p_i, k_{ij}) S_{-\lambda}(k_{ij}, p_{ij}, p_j, k_j) \nonumber \\
& + (v + \lambda a) m_i m_j S_{-\lambda}(k_i, p_{ij}, k_{ij}) S_{-\lambda}(k_{ij}, k_j) \bigg]
\end{aligned} \nonumber \\
M&^{V \rightarrow f\bar{f}}(\lambda, -\lambda, \lambda) = A^{\mbox{\tiny{split}}}_{\perp} \nonumber \\
\times &
\begin{aligned}[t]
& \bigg[ (v + \lambda a) S_{-\lambda}(k_i, p_i, p_{ij}, k_{ij}) \nonumber \\
& + (v - \lambda a) m_i m_j S_{-\lambda}(k_i, k_{ij}) S_{-\lambda}(k_{ij}, p_{ij}, k_j) \bigg]
\end{aligned} \nonumber \\
M&^{V \rightarrow f\bar{f}}(\lambda, \lambda, \lambda) = A^{\mbox{\tiny{split}}}_{\perp} \nonumber \\
\times &
\begin{aligned}[t]
& \bigg[ (v + \lambda a) m_i S_{-\lambda}(k_i, p_{ij}, k_{ij}) S_{-\lambda}(k_{ij}, p_j, k_j) \nonumber \\
& + (v - \lambda a) m_j S_{-\lambda} (k_i, p_i, k_{ij}) S_{-\lambda}(k_{ij}, p_{ij}, k_j) \bigg]
\end{aligned} \nonumber \\
M&^{V \rightarrow f\bar{f}}(\lambda, -\lambda, -\lambda) = A^{\mbox{\tiny{split}}}_{\perp} \nonumber \\
\times &
\begin{aligned}[t]
& \bigg[ (v - \lambda a) m_i S_{-\lambda}(k_i, k_{ij}) S_{-\lambda}(k_{ij}, p_{ij}, p_{j}, k_j) \nonumber \\
& + (v + \lambda a) m_j S_{-\lambda}(k_i, p_i, p_{ij}, k_{ij}) S_{-\lambda}(k_{ij}, k_j) \bigg]
\end{aligned} \nonumber \\
M&^{V \rightarrow f\bar{f}}(0, \lambda, -\lambda) = A^{\mbox{\tiny{split}}}_{L} \nonumber \\
\times &
\begin{aligned}[t]
& \bigg[ S_{-\lambda}(k_i, (v-\lambda a) (m_i^2 p_j + m_j^2 p_i) \nonumber \\
& - (v + \lambda a) m_i m_j (p_{ij} - \frac{m_{ij}^2}{p_{ij}{\cdot}k_{ij}}) k_{ij} , k_j ) \nonumber \\
& - \frac{m_{ij}^2}{p_{ij}{\cdot}k_{ij}} (v - \lambda a) S_{-\lambda}(k_i, p_i, k_{ij}, p_j, k_j) \bigg]
\end{aligned} \nonumber \\
M&^{V \rightarrow f\bar{f}}(0, \lambda, \lambda) = A^{\mbox{\tiny{split}}}_{L} \nonumber \\
\times &
\begin{aligned}[t]
& \bigg[ m_i (v + \lambda a) S_{-\lambda} (k_i, p_{ij} - \frac{m_{ij}^2}{p_{ij}{\cdot}k_{ij}}k_{ij}, p_j, k_j) \nonumber \\
& - m_j (v - \lambda a) S_{-\lambda}(k_i, p_{i}, p_{ij} - \frac{m_{ij}^2}{p_{ij}{\cdot}k_{ij}}, k_j) \bigg]
\end{aligned} \nonumber \\
\end{align}

\subsection{Higgs Splitting}
\subsubsection{Higgs Splitting to Fermion-antifermion}
\begin{align}
M&^{f \rightarrow f h}(\lambda, \lambda, h) = \frac{e}{4 s_w} \frac{m_i}{m_w} \frac{1}{\sqrt{p_{i}{\cdot}k_{i}} \sqrt{p_{j}{\cdot}k_j}} \nonumber \\
&\times \bigg[ S_{-\lambda}(k_i, p_i, p_j, k_j) - m_i^2 S_{-\lambda}(k_i, k_j) \bigg] \nonumber \\
M&^{f \rightarrow f h}(\lambda, -\lambda, h) = \frac{e}{4 s_w} \frac{m_i^2}{m_w} \frac{1}{\sqrt{p_{i}{\cdot}k_{i}} \sqrt{p_{j}{\cdot}k_j}} \nonumber \\
& \times S_{-\lambda}(k_i, p_i - p_j, k_j)
\end{align}

\subsubsection{Higgs Splitting to Vector Bosons}
\begin{align}
M&^{h\rightarrow VV}(h, \lambda, -\lambda) = -\frac{g_{h}}{4} \nonumber \\
\times & S_{-\lambda}(k_i, p_i, k_j) S_{-\lambda}(k_i, p_j, k_j) \nonumber \\
M&^{h\rightarrow VV}(h, \lambda, \lambda) = -\frac{g_{h}}{4} \nonumber \\
\times & S_{-\lambda}(k_j, k_i) S_{-\lambda}(k_i, p_i, p_j, k_j) \nonumber \\
M&^{h\rightarrow VV}(h, 0, \lambda) = - \frac{g_{h}}{2\sqrt{2}} \frac{1}{m_i} \frac{\lambda}{p_j{\cdot}k_j} \nonumber \\
\times & S_{-\lambda}(k_j, p_j, p_i - \frac{m_i^2}{p_i{\cdot}k_i} k_i, k_j) \nonumber \\
M&^{h\rightarrow VV}(h, \lambda, 0) - \frac{g_{h}}{2\sqrt{2}} \frac{1}{m_j} \frac{\lambda}{p_i{\cdot}k_i} \nonumber \\
\times &= S_{-\lambda}(k_i, p_i, p_j - \frac{m_j^2}{p_j{\cdot}k_k} k_j, k_i) \nonumber \\
M&^{h\rightarrow VV}(h, 0, 0) = \frac{g_{h}}{m_i m_j} \nonumber \\
\times & \bigg[\frac{1}{2}\left(m_{ij}^2 - m_i^2 - m_j^2 \right) - m_j^2 \frac{p_i{\cdot}k_i}{p_j{\cdot}k_j} - m_i^2 \frac{p_j{\cdot}k_j}{p_i{\cdot}k_i}\bigg]
\end{align}


\subsection{Vector Boson Emission (Initial State)}
\subsubsection{Vector Boson Emission from Fermion}
We define the prefactors 
\begin{align}
\tilde{A}^{\mbox{\tiny{emit}}}_{\perp} &= \frac{1}{2 \sqrt{2}} \frac{\lambda}{\sqrt{p_{a}{\cdot}k_{j}} \sqrt{p_{aj}{\cdot}k_{aj}} \, p_j{\cdot}k_j} \nonumber \\
\tilde{A}^{\mbox{\tiny{emit}}}_{L} &= \frac{1}{2} \frac{1}{m_j} \frac{1}{\sqrt{p_a {\cdot} k_a} \sqrt{p_{aj}{\cdot}k_{aj}}}.
\end{align}

\begin{align}
M&^{\tilde{f} \rightarrow \tilde{f}' V}(\lambda, \lambda, \lambda) = \tilde{A}^{\mbox{\tiny{emit}}}_{\perp} \nonumber \\
\times &
\begin{aligned}[t]
& \bigg[(v - \lambda a) S_{-\lambda}(k_{aj}, p_{aj}, p_j, k_j) S_{-\lambda}(k_j, p_a, k_a) \nonumber \\
& - (v + \lambda a) m_a m_{aj} S_{-\lambda}(k_{aj}, k_j) S_{-\lambda}(k_j, p_j, k_a) \bigg]
\end{aligned} \nonumber \\
M&^{\tilde{f} \rightarrow \tilde{f}' V}(\lambda, \lambda, -\lambda) = \tilde{A}^{\mbox{\tiny{emit}}}_{\perp} \nonumber \\
\times &
\begin{aligned}[t]
&\bigg[ (v - \lambda a) S_{-\lambda}(k_{aj}, p_{aj}, k_j) S_{-\lambda}(k_j, p_j, p_a, k_,a)\nonumber \\
& - (v + \lambda a) m_a m_{aj} S_{-\lambda}(k_{aj}, p_j, k_j) S_{-\lambda}(k_j, k_a) \bigg]
\end{aligned} \nonumber \\
M&^{\tilde{f} \rightarrow \tilde{f}' V}(\lambda, -\lambda, \lambda) = \tilde{A}^{\mbox{\tiny{emit}}}_{\perp} \nonumber \\
\times &
\begin{aligned}[t]
& \bigg[ (v + \lambda a) m_{aj} S_{-\lambda}(k_{aj}, k_j) S_{-\lambda}(k_j, p_j, p_a, k_a) \nonumber \\
& - (v - \lambda a) m_a S_{-\lambda}(k_{aj}, p_{aj}, p_j, k_j) S_{-\lambda}(k_j, k_a) \bigg]
\end{aligned} \nonumber \\
M&^{\tilde{f} \rightarrow \tilde{f}' V}(\lambda, -\lambda, -\lambda) = \tilde{A}^{\mbox{\tiny{emit}}}_{\perp} \nonumber \\
\times &
\begin{aligned}[t]
&\bigg[ (v + \lambda a) m_{aj} S_{-\lambda}(k_{aj}, p_j, k_j) S_{-\lambda}(k_j, p_a, k_a)\nonumber \\
& - (v - \lambda a) m_a S_{-\lambda}(k_{aj}, p_{aj}, k_j) S_{-\lambda}(k_j, p_j, k_a) \bigg]
\end{aligned} \nonumber \\
M&^{\tilde{f} \rightarrow \tilde{f}'V}(\lambda, \lambda, 0) = \tilde{A}^{\mbox{\tiny{emit}}}_{L} \nonumber \\
\times &
\begin{aligned}[t]
& \bigg[ S_{-\lambda}(k_{aj}, (v-\lambda a)(m_a^2 p_{aj} - m_{aj}^2 p_{a}) \nonumber \\
&+ (v+\lambda a)m_a m_{aj} p_j, k_{a}) \nonumber \\
& - \frac{m_j^2}{p_j{\cdot}k_j}  \bigg( (v-\lambda a) S_{-\lambda}(k_{aj}, p_{aj}, k_j, p_a, k_a) \nonumber \\
& - (v + \lambda a) m_{aj} m_a S_{-\lambda}(k_{aj}, k_j, k_{a}) \bigg) \bigg] \nonumber \\
\end{aligned} \nonumber \\
M&^{\tilde{f} \rightarrow \tilde{f}'V}(\lambda, -\lambda, 0) = \tilde{A}^{\mbox{\tiny{emit}}}_{L} \nonumber \\
\times &
\begin{aligned}[t]
& \bigg[ m_{aj} (v-\lambda a)  S_{-\lambda}(k_{aj}, p_j - \frac{m_j^2}{p_j{\cdot}k_j} k_j, p_{a}, k_{a}) \nonumber \\
& + m_{aj} (v+\lambda a) S_{-\lambda} ( k_{aj}, p_{aj}, p_j - \frac{m_j^2}{p_j{\cdot}k_j} k_j, k_{a}) \bigg]
\end{aligned} \nonumber \\
\end{align}

\subsubsection{Vector Boson Emission from Antifermion}
\begin{align}
M&^{\tilde{\bar{f}} \rightarrow \tilde{\bar{f}}' V}(\lambda, \lambda, \lambda) = \tilde{A}^{\mbox{\tiny{emit}}}_{\perp} \nonumber \\
\times &
\begin{aligned}[t]
& \bigg[(v + \lambda a) S_{\lambda}(k_{a}, p_{a}, k_j) S_{-\lambda}(k_j, p_j p_{aj}, k_{aj}) \nonumber \\
& - (v - \lambda a) m_a m_{aj} S_{\lambda}(k_{a}, p_j, k_j) S_{-\lambda}(k_j, k_{aj}) \bigg]
\end{aligned} \nonumber \\
M&^{\tilde{\bar{f}} \rightarrow \tilde{\bar{f}}' V}(\lambda, \lambda, -\lambda) = \tilde{A}^{\mbox{\tiny{emit}}}_{\perp} \nonumber \\
\times &
\begin{aligned}[t]
&\bigg[ (v + \lambda a) S_{\lambda}(k_{a}, p_{a}, p_j, k_j) S_{-\lambda}(k_j, p_{aj}, k_{aj})\nonumber \\
& - (v - \lambda a) m_a m_{aj} S_{\lambda}(k_{a}, k_j) S_{-\lambda}(k_j, p_j, k_a) \bigg]
\end{aligned} \nonumber \\
M&^{\tilde{\bar{f}} \rightarrow \tilde{\bar{f}}' V}(\lambda, -\lambda, \lambda) = \tilde{A}^{\mbox{\tiny{emit}}}_{\perp} \nonumber \\
\times &
\begin{aligned}[t]
& \bigg[ (v - \lambda a) m_a S_{\lambda}(k_{a}, p_j, k_j) S_{-\lambda}(k_j, p_{aj}, k_{aj}) \nonumber \\
& - (v + \lambda a) m_{aj} S_{\lambda}(k_{a}, p_{a}, k_j) S_{-\lambda}(k_j, p_j, k_{aj}) \bigg]
\end{aligned} \nonumber \\
M&^{\tilde{\bar{f}} \rightarrow \tilde{\bar{f}}' V}(\lambda, -\lambda, -\lambda) = \tilde{A}^{\mbox{\tiny{emit}}}_{\perp} \nonumber \\
\times &
\begin{aligned}[t]
& \bigg[ (v - \lambda a) m_a S_{\lambda}(k_{a}, k_j) S_{-\lambda}(k_j, p_j, p_{aj}, k_{aj})\nonumber \\
& - (v + \lambda a) m_{aj} S_{\lambda}(k_{a}, p_{a}, p_j, k_j) S_{-\lambda}(k_j, k_{aj}) \bigg]
\end{aligned} \nonumber \\
M&^{\tilde{\bar{f}} \rightarrow \tilde{\bar{f}}' V}(\lambda, \lambda, 0) = \tilde{A}^{\mbox{\tiny{emit}}}_{L} \nonumber \\
\times &
\begin{aligned}[t]
&\bigg[ S_{\lambda}(k_{a}, (v+\lambda a)(m_a^2 p_{aj} - m_{aj}^2 p_{a}) \nonumber \\
&+ (v-\lambda a)m_a m_{aj} p_j, k_{aj}) \nonumber \\
& - \frac{m_j^2}{p_j{\cdot}k_j}  \bigg( (v+\lambda a) S_{\lambda}(k_{a}, p_{a}, k_j, p_{aj}, k_{aj}) \nonumber \\
&- (v - \lambda a) m_{aj} m_a S_{\lambda}(k_{a}, k_j, k_{aj}) \bigg) \bigg]
\end{aligned} \nonumber \\
M&^{\tilde{\bar{f}} \rightarrow \tilde{\bar{f}}' V}(\lambda, -\lambda, 0) = \tilde{A}^{\mbox{\tiny{emit}}}_{L} \nonumber \\
\times &
\begin{aligned}[t]
& \bigg[ m_a (v-\lambda a)  S_{\lambda}(k_{a}, p_j - \frac{m_j^2}{p_j{\cdot}k_j} k_j, p_{aj}, k_{aj}) \nonumber \\
&  + m_{aj} (v+\lambda a) S_{\lambda} ( k_{a}, p_{a}, p_j - \frac{m_j^2}{p_j{\cdot}k_j} k_j, k_{aj}) \bigg]
\end{aligned} \nonumber \\
\end{align}

\bibliographystyle{spphys}
\bibliography{ref} 
\end{document}